\documentclass{article}

\usepackage[numbers]{natbib}
\usepackage[preprint]{neurips}

\usepackage{makecell}
\usepackage{graphicx}
\usepackage{subcaption}

\usepackage{amsmath}
\usepackage{mathtools}
\usepackage{bm}
\usepackage{svg}

\usepackage[linesnumbered,ruled,vlined]{algorithm2e}

\usepackage[utf8]{inputenc} % allow utf-8 input
\usepackage[T1]{fontenc}    % use 8-bit T1 fonts
\usepackage{hyperref}       % hyperlinks
\usepackage{url}            % simple URL typesetting
\usepackage{booktabs}       % professional-quality tables
\usepackage{amsfonts}       % blackboard math symbols
\usepackage{nicefrac}       % compact symbols for 1/2, etc.
\usepackage{microtype}      % microtypography
\usepackage{xcolor}         % colors

\usepackage{tikz}
\usepackage{array}

\usepackage{caption} % <--- important
\title{A Hierarchical Adaptive Diffusion Model for Flexible Protein--Protein Docking}

\author{%
  Rujie Yin\textsuperscript{1} \quad Yang Shen\textsuperscript{1} \\
  \textsuperscript{1}Department of Electrical and Computer Engineering, Texas A\&M University \\
  \texttt{\{rujieyin, yshen\}@tamu.edu}
}

\begin{document}

\maketitle

\begin{abstract}
%   The abstract paragraph should be indented \nicefrac{1}{2}~inch (3~picas) on
%   both the left- and right-hand margins. Use 10~point type, with a vertical
%   spacing (leading) of 11~points.  The word \textbf{Abstract} must be centered,
%   bold, and in point size 12. Two line spaces precede the abstract. The abstract
%   must be limited to one paragraph.
Structural prediction of protein--protein interactions is important to understand molecular basis of cellular interactions.  However, it still faces major challenges especially when significant conformational changes are present.  We propose a generative framework of hierarchical adaptive diffusion to improve accuracy and efficiency in such cases. It is \textit{hierarchical} in separating global inter-protein rigid-body motions and local intra-protein flexibility in diffusion processes; and the distinct local and global noise schedules are designed to mimic the induced fit effect.  It is \textit{adaptive} in conditioning the local flexibility schedule on predicted levels of conformational changes, allowing for faster flexing for larger anticipated conformational changes.  Furthermore, it couples the local and global diffusion processes through a common score and confidence network with sequence, evolution, structure, and dynamics features as inputs and maintains rotational or translational invariance/equivariance in outputs.  Lastly, it builds on our newly curated DIPS-AF dataset of nearly 39,000 examples for pre-training.  Numerical experiments on independent docking benchmark dataset DB5.5 show that our model outperforms an AlphaFold2-like iterative transformer (GeoDock) and a diffusion model (DiffDock-PP) in both rigid and flexible cases, with larger improvement margins in more flexible cases.  Ablation studies prove the importance of adaptive schedules, dynamics features, and pre-training.  Additional analyses and case studies reveal remaining gaps in sampling, scoring, and resolution of conformations.   
\end{abstract}

\section{Introduction}

Protein--protein interactions underlie many essential biological processes~\cite{alberts2022molecular}. Accurate modeling of these interactions, including their 3-dimensional (3D) structural basis,  is critical for understanding cellular function and for facilitating drug discovery \cite{vakser2014protein,nooren2003diversity}. However, predicting the 3D structure of a protein--protein  complex remains challenging in the post-AlphaFold2 era, particularly when substantial conformational changes are involved  \cite{harmalkar2025reliable}. 

Traditional protein docking methods sample conformational space following physical or statistical scoring functions. Given the enormously high-dimensional conformational space and the extremely noisy scoring functions, they 
either rely on rigid‐body approximations~\cite{zdock_ref,cluspro_ref} for speed or incorporate backbone and side-chain flexibility through enhanced sampling techniques~\cite{may2005accounting, wang2007protein,marze2018efficient,harmalkar2021advances,christoffer2022domain,harmalkar2025reliable}.
Whereas rigid docking ignores conformational changes critical for protein--protein interactions and protein functions~\cite{desta2020performance}, flexible docking can be computationally demanding and challenging.  

% or by docking against an ensemble of pre-generated conformations which can be accurate, albeit at the cost of forbiddenly large computational resources and processing time \cite{may2005accounting, wang2007protein, desta2020performance}. 

Recent advances in deep learning methods for protein docking, including diffusion‐based generative models, learn the geometry (distributions) of 3D protein complex structures directly from data.  They have the potential to improve the accuracy and efficiency of protein docking. 
However, they face the same challenge of high-dimensional conformational space as traditional docking methods did. In practice, they again either assume conformational  rigidity~\cite{diffdock_pp_ref} or allow certain conformational flexibility at an enormous computational cost~\cite{abramson2024accurate}.

% Nevertheless, many such approaches still either restrict flexibility to a few key degrees of freedom assuming a rigid-docking scenario or attempt to model full atomic flexibility via high-dimensional point-cloud representations. In the former case, such simplifications may fail to capture essential induced-fit conformational rearrangements, while in the latter case, the inherent rigidity present may be overlooked \cite{diffdock_pp_ref, geodock_ref}.

\begin{figure}[htbp]
    \centering
    \includegraphics[width=\linewidth]{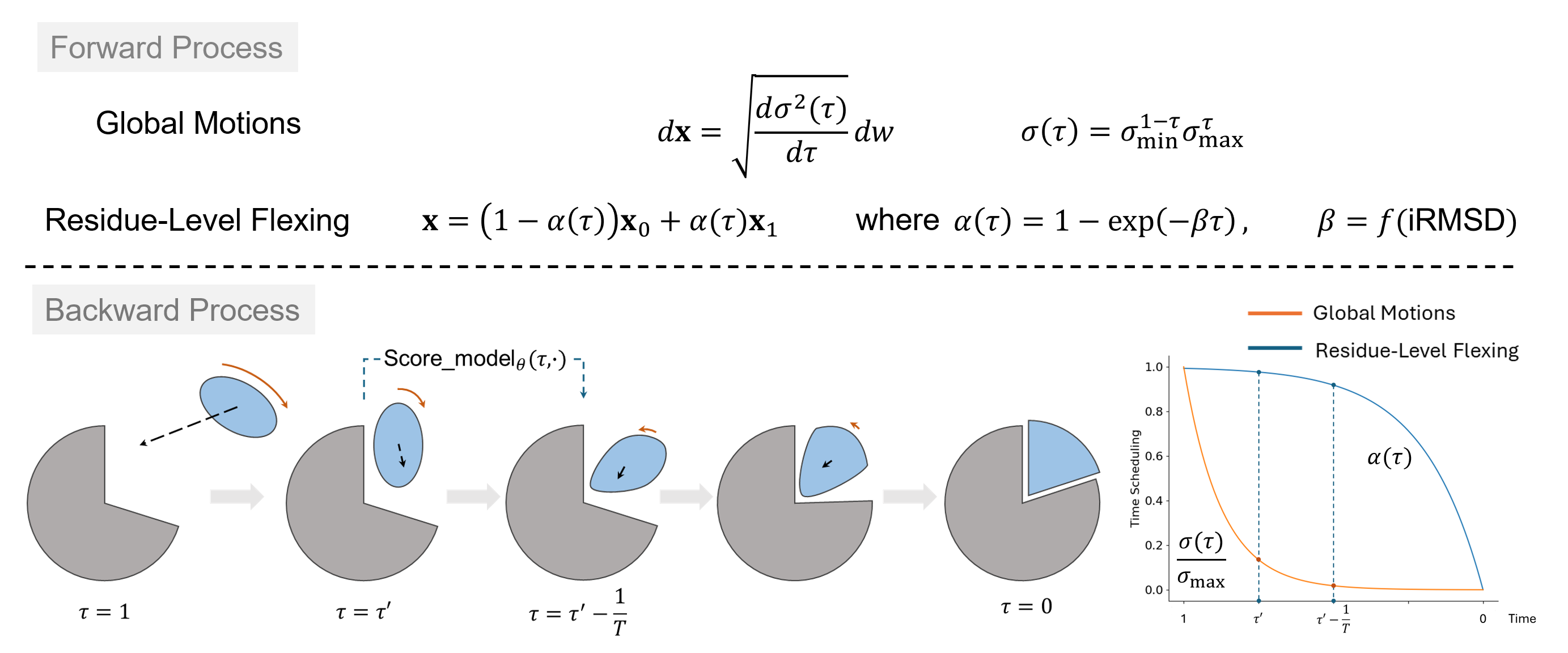}
    % \includesvg[width=\linewidth]{Docking_Scheme_3.svg}
    \caption{\textbf{Illustration of the hierarchical adaptive diffusion process on the product space \(\mathbb{P} = \mathbb{R}^3 \times SO(3) \times (\mathbb{R}^3 \times SO(3))^{n_1 + n_2}\), capturing both global motions and residue-level flexing.} In the above portion, the diffusion variable \(\mathbf{x}\) in the forward process represents both translational and rotational components for its corresponding level of transformation. For the bottom left figure, the backward/sampling process guided by update signal derived from outputs of score model evolves from an initial unbound state toward the final docked conformation, incorporating (i) global rigid-body translation and rotation of the ligand and (ii) residue-level translations and rotations for fine-grained local flexibility. For the bottom right figure, it shows the time scheduling for global protein-level and local residue-level diffusion processes.}
    \label{fig:overall_illustration_new}
\end{figure}

In this work, we propose a novel, generative protein-docking framework through hierarchical adaptive diffusion (Figure~\ref{fig:overall_illustration_new}).  First, we parameterize the conformational space hierarchically as global protein-level rigid-body motions and local residue-level flexibility.  Second, we model global and local translation and rotation diffusion processes with distinctive noise schedules, increasing local flexibility as proteins approach each other in global poses to mimic the induced fit effect.  We further employ an adaptive flexibility schedule that conditions on predicted conformational changes and starts earlier in reverse diffusion with higher levels of predicted conformational changes.  Third, we couple the local and global diffusion processes through a common score model that utilizes sequence, evolution, structure, and dynamics features and maintains desired rotational/translational invariance/equivariance for various outputs.  Last, we curate a DIPS-AF dataset of nearly 39,000 protein--protein pairs and pretrain our models for better accuracy and generalizability.  Numerical performances on an independent benchmark set DB5.5 show that our model outperforms GeoDock (iterative transformer for flexible docking) and DiffDock-PP (diffusion for rigid docking) in all levels of docking difficulty, especially when conformational changes are significant.

We next detail a survey of relevant protein docking methods, traditional docking or recent learning, as well as remaining gaps and our contributions in Sec.~\ref{sec:related-methods}. 

\section{Related Work and Our Contributions}\label{sec:related-methods}
Protein--protein docking methods can be broadly classified into two categories: traditional docking models and deep learning based docking models.

\textbf{Traditional Docking Models.}
Traditional protein-docking methods, such as ClusPro~\cite{cluspro_ref}, ZDOCK~\cite{zdock_ref}, HADDOCK~\cite{haddock_ref}, and RosettaDock~\cite{rosettadock_ref}, have been widely used. These methods typically employ exhaustive sampling techniques—using FFT-based algorithms or Monte Carlo simulations—to explore the rigid-body orientation space, followed by energy-based scoring and clustering to select the best candidates. They are often parameterized under a rigid-body assumption (e.g., ClusPro, ZDOCK) or incorporate limited flexibility (e.g., HADDOCK, RosettaDock) by allowing adjustments in side-chains or local backbone segments. Although these methods are robust for many targets, their assumption of minimal conformational change upon binding restricts their applicability to complexes with significant induced fit. Simulation-based methods, such as Monte-Carlo simulations and molecular dynamics~\cite{replicadock_ref}, although allowing for atom-level conformational flexibility, remain computationally demanding.

% Furthermore, the computational cost associated with running a full simulation for even a single target remains high, making large-scale batch processing challenging.

\textbf{Deep Learning Based Docking Models.}
Deep learning approaches have revolutionized protein--protein docking by leveraging neural architectures to learn complex interactions from structural and sequence data. Unlike traditional physics- or statistics-driven methods, these models directly infer docking configurations from large-scale datasets, often guided by geometric representations or generative modeling techniques.  These models are typically subdivided into rigid and flexible categories based on how they treat deformation parametrization:

\emph{Rigid-Body Docking Models.}  
Rigid-body docking methods such as DiffDock-PP~\cite{diffdock_pp_ref}, EquiDock~\cite{equidock_ref}, dMaSIF-based docking~\cite{dmasif_dock_ref}, and BiDock (xTrimoBiDock)~\cite{bidock_ref} predict the relative orientation of docking partners using learned geometric representations. These models typically leveraging geometric neural networks and/or diffusion-based generative frameworks. They are highly efficient and often outperform traditional methods on targets with near-bound conformations. Particularly, generative model based approaches enable continuous sampling in the pose space, allowing for uncertainty estimation and the generation of diverse docking hypotheses, which is advantageous in capturing alternative binding modes~\cite{diffdock_pp_ref}. However, by assuming rigidity, these methods cannot account for the substantial conformational adjustments that may be required for accurate docking.

\emph{Flexible Docking Models.}  
To address the shortcomings of rigid-body approaches, several flexible docking models have emerged in the recent years. Notable among these are AlphaFold-Multimer~\cite{alphafold_multimer_ref}, GeoDock~\cite{geodock_ref}, and  DockGPT~\cite{dockgpt_ref}. 
% , and ReplicaDock 2.0 (and related hybrid pipelines)~\cite{replicadock_ref}. 
AlphaFold-Multimer extends AlphaFold 2 to multi-chain protein structure prediction through updates in inputs, model architecture, and structure losses~\cite{alphafold_multimer_ref}.  GeoDock employs a transformer-based iterative refinement strategy similar to AlphaFold 2. Given binding pocket information, DockGPT uses a generative transformer to sample multiple docked complexes, effectively incorporating flexibility into the docking process. 
%AlphaFold-Multimer predicts complex structures directly from sequence information (co-folding), implicitly modeling conformational changes. %, while ReplicaDock 2.0 integrates physics-based Monte Carlo sampling with backbone flexibility refinement. Although these methods significantly improve accuracy on targets with induced fit, they often lack a systematic mechanism to balance global and local flexibility.

\textbf{Gaps and Contributions.}  
Despite these advances, current methods face two major challenges. First, rigid-body approaches fail to capture the conformational rearrangements inherent in many protein--protein interactions. Second, flexible docking methods, while more accurate on highly flexible targets, typically sample  flexibility uniformly for each case or rely on ad hoc refinement steps, leading to computational inefficiencies and inconsistent performance. 

Our proposed hierarchical adaptive docking model addresses these gaps by combining a multi-scale representation with an adaptive flexing schedule parametrized by predicted iRMSD. Additionally, we integrate Normal Mode-derived features to further refine residue-level flexibility predictions, leveraging intrinsic protein dynamics. Our key contributions are as follows (shown in Figure~\ref{fig:overall_illustration_new}):
\begin{itemize}
    \item \emph{Hierarchical Representation.} We introduce a multi-scale framework that represents the protein complex as the product of transformation groups---accounting for global inter-protein  motions (both translation and rotation) and local  intra-protein flexibility (residue-frame rotations and translations). This formulation treats both global and local transformations in a homogeneous space, allowing the model to efficiently capture large-scale domain movements as well as fine-grained local adjustments.
    \item \emph{Distinctive Global and Local Diffusion Schedules for Induce Fit.}  We design different noise schedules for global and local diffusion processes to mimic induced fit effects --- as the reverse denoising diffusion progresses, it shifts its focus from global inter-protein rigid-body motions to local intra-protein flexibility.  Moreover,  rather than applying the same fixed flexing schedule for all proteins, we introduce an adaptive flexing schedule conditioned on a predicted deformation metric (interface RMSD). This adaptive scheduling can enable local flexibility  earlier in the reverse diffusion process if larger conformational changes are anticipated for given protein pairs.
    \item \emph{Graph Representations with Sequence, Evolution, Structure, and Dynamics Features.}  We represent 3D structures of protein complexes as intra and inter-protein graphs with nodes and edges featuring protein sequence, evolution (from a protein language model), structure, and dynamics (from normal mode analysis).  Particularly, the newly-introduced dynamics features provide biologically meaningful descriptors of intrinsic protein flexibility and help guide the docking process by highlighting regions that are prone to significant conformational changes upon binding.
    \item \emph{Score and Confidence Model with $SE(3)$ Invariance/Equivariance}. We couple local and global rotational and translational diffusions through a common score and confidence model.  With the graph representations above as the input, this model outputs local and global translational and rotational scores as well as quality (confidence) estimates of clddt values.  We design a tensor product interaction network and constrain spherical harmonics of various degrees for outputs of various rototranslational invariance or equivariance needs.
    \item \emph{Pre-training with curated DIPS-AF dataset}.   To overcome the data scarcity challenge in current protein-docking benchmark sets (DB5.5 contains hundreds of cases), we curated a DIPS-AF dataset of nearly 39,000 pairs of unbound and bound protein--protein structures and used it to pretrain and improve our diffusion model.
\end{itemize}

% \section{Preliminaries}
% \subsection{Normal Modes Analysis}

% Normal Modes Analysis (NMA) is a powerful tool to study the intrinsic flexibility of macromolecules under the harmonic approximation. Starting from an equilibrium conformation \(\mathbf{x}_0\), the potential energy is locally approximated by
% \[
% V(\mathbf{x}) \approx V(\mathbf{x}_0) + \frac{1}{2}(\mathbf{x}-\mathbf{x}_0)^T H (\mathbf{x}-\mathbf{x}_0),
% \]
% where \(H = \nabla^2 V(\mathbf{x}_0)\) is the Hessian matrix. The normal modes are obtained by solving the eigenvalue problem
% \[
% H\,\mathbf{q}_i = \lambda_i\,\mathbf{q}_i,
% \]
% with eigenvalues \(\lambda_i\) representing the stiffness along the corresponding eigenvectors \(\mathbf{q}_i\). The system's displacement can then be expressed as a linear combination of these modes,
% \[
% \Delta \mathbf{x} = \sum_{i} a_i \,\mathbf{q}_i,
% \]
% and, the mean square fluctuation of a residue \(j\) is approximated by
% \[
% \text{MSF}_j \propto \sum_i \frac{(q_{ij})^2}{\lambda_i},
% \]
% where \(q_{ij}\) is the \(j\)th component of \(\mathbf{q}_i\). Typically, only the first few nontrivial modes (i.e., those with the smallest nonzero eigenvalues) are considered, as they capture the most significant, biologically relevant motions.

\section{Method}

% \subsection{An effective parametrization of a-priori \textit{plausible} protein space}
\subsection{Parametrization of the Conformational Space for Protein Docking}
There is a spectrum of methods parametrizing the flexible protein-protein docking problem. On the two ends of the spectrum, they are simplification of the problem into the rigid body docking problem and relaxation of the problem into fully flexible docking only dealing with limited rigidity resulting from a set of admissible torsions. The two ends of the formulation, respectively, should incur oversimplification trouble by ignoring the intra-protein flexibility totally and model overparametrization problem caused by not considering some inherent rigidity.  The solution to the above two problems is to find a proper spot in the parametrization spectrum.\\

% The illustration of the two types of models is as shown in Table~\ref{docking spectrum model performance table}.
% \begin{table}
%     \caption{\textbf{Docking Spectrum on DB5.5.} Performance comparison between DiffDock-PP and  GeoDock to represent methods residing on the two ends of the sprectrum. Both models are trained on the DIPS training set and tested on the DB5.5 test set. The results Complex-RMSD and Interface-RMSD in \AA $ $ are reported for the 25 DB5.5 test complexes. (*GeoDock only successfully evaluated 24 out of 25 complexes; Evaluation of the complex 1N2C exceeded the  memory constraint for an 80 GB GPU.)}
% \label{docking spectrum model performance table}
%   \centering
%   \begin{tabular}{lll|ll}
%     \toprule
%     % \multicolumn{2}{c}{Part}                   \\
%     % \cmidrule(r){1-3}
%     & \multicolumn{2}{c}{\textbf{Complex-RMSD (Å)}} 
%     & \multicolumn{2}{c}{\textbf{Interface-RMSD (Å)}} \\
%     \cmidrule(lr){2-3} \cmidrule(lr){4-5}
%     \textbf{Methods} & Mean & Median & Mean & Median \\
%     \midrule
%     DiffDock-PP (fully rigid)     &   16.28 & 16.52  &   16.04 & 16.23   \\
%     GeoDock*  (fully flexible)   &     15.51   &  15.62   &   14.84    &     13.83  \\
%     %     &       &   \\
%     \bottomrule
%   \end{tabular}
% \end{table}

\subsubsection{One end: full flexibility as flat parameterizations}
Fully flexible protein-protein docking can typically be formulated as determining the  complex structure parametrized by the backbone frames' positions and orientations (such as AlphaFold 2 and GeoDock~\cite{geodock_ref}),  the atom coordinates (such as AlphaFold 3~\cite{abramson2024accurate}), or the internal coordinates of torsions.  These ``flat'' parameterizations do not distinguish global and local conformational variables either in parameterization or sampling process.    
% (and their associated torsion angles if the model takes full-atom representation). In this formulation, the complex resides on a submanifold \(\mathcal{M}_{com} \subset \mathbb{R}^{6(n_1+n_2)},\) where \(n_1\) and \(n_2\) denote the numbers of residues (i.e., backbone frames) in each protein. This full-flexibility point-cloud-based parametrization, however, suffers from some inherent rigidity not explicitly taken care of, often resulting in poor performance in backbone conformation modeling; GeoDock model is a representative method that illustrates the challenges.

\subsubsection{The other end: rigid-body assumption}
On the other hand, flexible protein-protein docking problem can also be formulated as determining the ligand pose in the space of a $6$-dimensional manifold $\mathbb{R}^3\times SO(3)$. The intra-flexibility for each protein is ignored in this case. This kind of method, however, also perform poorly when they predict the formed complex structures from two unbound protein structures. DiffDock-PP model is used to illustrate this end.

\subsubsection{A balanced approach: hierarchical modeling of global and local flexibility}

This parameterization introduces a hierarchical framework for protein-protein docking that integrates both global and local motions within a unified mathematical representation in \(\mathbb{R}^3 \times SO(3)\) spaces.

The \(((n_1 + n_2)\times 6 + 6)\)-dimensional manifold \(\mathcal{M}_{com}\), where the receptor-ligand complex backbone structure resides, is constructed as a series of transformations. These transformations include rigid-body translation and rotation of the ligand, as well as residue-level intra-conformational flexibility for both receptor and ligand. The residue-level transformations are parameterized by backbone frame positions (\(C\alpha\) positions) \(\mathbf{x}\) and frame orientations \(O\) representing the orientations of the backbone atom \(N,C\alpha,C\) triangles. These transformations allow fine-grained intra-flexibility while preserving structural consistency.

The overall manifold \(\mathcal{M}_{com}\) is mapped to the product space of these transformation groups:
\[
\mathbb{P} = \mathbb{R}^3 \times SO(3) \times (\mathbb{R}^3 \times SO(3))^{n_1 + n_2}
\]
which accounts for rigid-body motions and residue-level intra-flexibility for both receptor and ligand. Here, \(\mathbb{R}^3\) denotes the translation space and \(SO(3)\) the rotation group.

The transformation groups act on an unbound complex structure \(\mathbf{c^*} \in \mathbb{R}^{(n_1 + n_2)\times 6+6}\), where \(\mathbf{x} = (\mathbf{x}_r, \mathbf{x}_l)\) denotes the receptor and ligand \textbf{\(C\alpha\) coordinates}, and \(O = (O_r, O_l)\) denotes their \textbf{backbone frames orientations}. The \textbf{global translation and global rotation} are denoted as \(\mathbf{t} \in \mathbb{R}^3\) and \(R \in SO(3)\), respectively, while \(\{(\mathbf{t}_{i_l}, R_{i_l})\} \in (\mathbb{R}^3 \times SO(3))^{n_l}\) are the set of \textbf{residue-level translation and rotation} pairs for each of the ligand residues \(j_r\), and \(\{(\mathbf{t}_{j_r}, R_{j_r})\} \in (\mathbb{R}^3 \times SO(3))^{n_r}\) are the set of \textbf{residue-level translation and rotation} pairs for receptor residues \(i_l\). 

Specifically, we have

Rigid-body translation:
\[
A_{tr}(\mathbf{t}, ((\mathbf{x}_r, O_r), (\mathbf{x}_l, O_l))) = ((\mathbf{x}_r, O_r), (\mathbf{x}_l + \mathbf{t}, O_l)),
\]

Rigid-body rotation:
\[
A_{rot}(R, ((\mathbf{x}_r, O_r), (\mathbf{x}_l, O_l))) = ((\mathbf{x}_r, O_r), (R(\mathbf{x}_l - \Bar{\mathbf{x}}_l) + \Bar{\mathbf{x}}_l, RO_l)),
\]
where \(\Bar{\mathbf{x}}_l\) is the center of mass of the ligand.

Residue-level transformations:
\[
A_\mathrm{res}(\{(\mathbf{t}_{j_r}, R_{j_r})\}, \{(\mathbf{t}_{i_l}, R_{i_l})\}, ((\mathbf{x}_r, O_r), (\mathbf{x}_l, O_l))) = 
\big(\{(\mathbf{x}_{j_r} + \mathbf{t}_{j_r}, R_{j_r} O_{j_r})\}, \{(\mathbf{x}_{i_l} + \mathbf{t}_{i_l}, R_{i_l} O_{i_l})\}\big),
\]

Composite transformation:
\[
A((\mathbf{t}, R, \{(\mathbf{t}_{j_r}, R_{j_r})\}, \{(\mathbf{t}_{i_l}, R_{i_l})\}), (\mathbf{x},O)) = 
A_{tr}(\mathbf{t}, A_{rot}(R, A_\mathrm{res}(\{(\mathbf{t}_{j_r}, R_{j_r})\}, \{(\mathbf{t}_{i_l}, R_{i_l})\}, (\mathbf{x},O)))).
\]

In each step of the sampling process, the updated ligand conformation resulted from intra-flexibility transformations is further constrained to the original ligand conformation in the RMSD sense using Kabsch algorithm, before rigid-body transformations are applied to it. This ensures that residue-level transformations are independent of rigid-body contributions, since residue-level translations are orthogonal to the global transformations after application of the Kabsch algorithm, while residue-level rotations are already disentangled with the global transformations applicable to frame positions defined by \(C\alpha\) atom positions only. \\

Finally, the space of complex structures is defined as:
\[
\mathcal{M}_{com} = \{A((\mathbf{t}, R, \{(\mathbf{t}_{j_r}, R_{j_r})\}, \{(\mathbf{t}_{i_l}, R_{i_l})\}), \mathbf{c^*}) \mid (\mathbf{t}, R, \{(\mathbf{t}_{j_r}, R_{j_r})\}, \{(\mathbf{t}_{i_l}, R_{i_l})\}) \in \mathbb{P}\}.
\]

This formulation explicitly models receptor and ligand residue positions and orientations, capturing both global and local transformations in a disentangled and physically consistent manner. Compared to the two extremes of the representation spectrum, our hierarchical representation adopts more realistic assumptions than the fully rigid model while also maintaining the overall docking position and orientation more explicitly than the fully flexible model in the homogeneous spaces (Translation and Rotation groups) between global protein-level and local residue levels. Additionally, the homogeneity of our parameterized space eliminates the lever effect, which poses feasibility challenges in integrating global transformations with intra-protein flexibility modeled through internal dihedral angles.

\subsection{Diffusion Process on the Product Space of Global and Local DoFs}
To construct a diffusion model on \(\mathbb{P}_\mathrm{res} = \mathbb{R}^3 \times SO(3) \times (\mathbb{R}^3 \times SO(3))^{n_1 + n_2}\), one only needs to devise a method capable of sampling from the SDE/ODE kernel on \(\mathbb{P}_\mathrm{res}\). Additionally, given that \(\mathbb{P}_\mathrm{res}\) is a product manifold, the forward diffusion process occurs separately within each constituent manifold. 

Therefore, it is sufficient to individually sample from each kernel and perform regression against its score within each subgroup, which is detailed in the next two subsections. 

\subsubsection{Variance Exploding SDE for Global Protein-Level Rigid-Body Motions}

In the four groups, for inter-protein poses, specifically ligand global rotation and translation while fixing the orientation and position of the receptor),  we use VE (Variance Exploding) SDE: \(\mathrm{d}x = \sqrt{\frac{\mathrm{d}\sigma^2(\tau)}{\mathrm{d}\tau}} \mathrm{d}w\) where \(\sigma^2 = \sigma^2_{\text{tr}}\), \( \sigma^2_{\text{rot}}\) respectively and \(w\) is the corresponding Brownian motion. For the rotation case, we diffuse on the isotropic Gaussian distribution on \(SO(3)\) (IG), as introduced by \cite{savjolova1985preface}, denoted as \(g \sim \text{IG}_{SO(3)}(\mu, \epsilon^2)\), is characterized by a mean rotation \(\mu\) and a scalar variance \(\epsilon\). This distribution can be expressed in an axis-angle format, where the axes are uniformly sampled and the rotation angle \(\omega\) ranges within \([0, \pi]\), presenting a density function

\[f(\omega) = \frac{1 - \cos \omega}{\pi} \sum_{l=0}^{\infty} (2l + 1)e^{-l(l+1)\epsilon^2} \frac{\sin\left((l + \frac{1}{2})\omega\right)}{\sin(\omega/2)}.\]

The uniform distribution on \(SO(3)\), denoted as \(U_{SO(3)}\), is specified with a uniform axis and 

\[f(\omega) = \frac{1-\cos \omega}{\pi},\]

which must be incorporated as a scaling factor when sampling from the distribution \cite{leach2022denoising}. \\

\subsubsection{Flow Matching with Adaptive Schedule for Local Residue-Level Flexibility}

For intra-protein flexibility of the receptor and ligand, we use a flow-matching model with novel adaptive schedule to model the residue frames' rotations and translations. 
% \subsection{Residual Flow Matching with Adaptive Residue-Level Flexibility Schedule}
% \subsection{Residual Flow Matching for Finer Flexibility in Protein Docking}

\textbf{Flow matching}. To efficiently capture the intrinsic residue-level flexibility required in protein docking, we use a residue-style-parametrized flow matching model that simulates the evolution of individual residue translations and rotations in terms of their residual corrections relative to the target conformation. In our formulation, the experimentally determined target conformation is assigned to time \(\tau=0\), while a less-structured prior conformation corresponds to \(\tau=1\) following the diffusion model notation. \\

Concretely, the dynamic of the residue-level rotations and translations are described below. Please note that we use superscript \(u\) to denote the unbound structures (respectively superimposed into their bound structures) and we omit the time \(\tau\) notation for simplicity.

For receptor residues, the residue-level translation and rotation of the \(j_r\)-th residue are defined as
\[
\begin{aligned}
\mathbf{t}_{j_r} &= \mathbf{x}_{j_r} - \mathbf{x}_{j_r}^u,\\[1ex]
R_{j_r} &= O_{j_r}(O_{j_r}^u)^{-1}.
\end{aligned}
\] 

For ligand residues - subject to global rotations \(R\) - the residue-level translation and rotation of the \(i_l\)-th residue are defined as
\[
\begin{aligned}
\mathbf{t}_{i_l} &= \mathbf{x}_{i_l} - \mathbf{x}_{i_l}^u(R),\\[1ex]
R_{i_l} &= O_{i_l}(O_{i_l}^u(R))^{-1}.
\end{aligned}
\] 

where the rotated unbound structures is
\(
\mathbf{x}_{i_l}^u(R) = A_{\text{rot}}(R, \mathbf{x}_{i_l}^u) = R\bigl(\mathbf{x}_{i_l}^u - \bar{\mathbf{x}}_l^u\bigr) + \bar{\mathbf{x}}_l^u,
\)
and the rotated unbound structures orientation is \(O_{i_l}^u(R) = A_{rot}(R, O_{i_l}^u) = RO_{i_l}^u\). \\

Let \(\mathbf{t}^\tau_\mathrm{res}\) (with rotations \(R^\tau_\mathrm{res}\)) denote the target residue-level translations (and rotations) at time \(\tau\) for either receptor or ligand. We define the intermediate state at time \(\tau\) as a linear interpolation,
\[
\mathbf{t}^\tau_\mathrm{res} = (1-\alpha(\tau))\,\mathbf{t}^0_\mathrm{res} + \alpha(\tau)\,\mathbf{t}^1_\mathrm{res},
\]
\[
R^\tau_\mathrm{res} = (1-\alpha(\tau))\,R^0_\mathrm{res} + \alpha(\tau)\,R^1_\mathrm{res},
\]
with the interpolation for rotations understood in an appropriate geodesic sense on \(SO(3)\).

\textbf{Adaptive schedule}. We employ a time schedule
\[
\alpha(\tau) = 1 - \exp(-\beta \tau),
\]
with \(\beta>0\) controlling the rate of conformational change. Notably, this schedule is designed so that as \(t\) approaches 0, the scaling factor \(\frac{\alpha'(\tau)}{1-\alpha(\tau)}\) increases significantly. This behavior mirrors the docking process where, upon contact, proteins exhibit more adjustments in their conformations. Additionally, rather than directly parameterizing the target conformation, our residual flow matching framework learns the residual correction \(\mathbf{x}_0 - \mathbf{x}_\tau\) and \(O_0(O_\tau)^{-1}\) (for both translations and rotations). That is, given an intermediate state \(\mathbf{x}_\tau, O_\tau\), a score model $s_\theta((\mathbf{x},O), \tau)$ is trained to predict the transformation between the target state and the current state, which are parametrized by the global protein-level and local residue-level transformations. Since in every step, we can predict the completely-flexed receptor and ligand proteins, we apply an additional interface-focused FAPE loss (as in the AlphaFold model) to further regulate the flexing prediction. More about the sampling process and loss is in Appendix~\ref{sec:appendixC}.\\

Furthermore, when the parameter \(\beta\) in the time schedule is larger, the flexing trajectory takes longer to complete, which corresponds to a greater conformational change and, consequently, a larger interface-RMSD (iRMSD) between the optimally docked unbound structures and the ground truth bound structures. We assume that when flexing is nearly complete (e.g., at 99\% of the process), the global translation should have progressed by a displacement equal to the iRMSD. Given that the noise schedule for global translation follows the geometric interpolation \( \sigma_{tr}(t) = \sigma_{tr\_min}^{\,1-t} \sigma_{tr\_max}^{\,t},
\) we define \(\beta\) as a function of the predicted iRMSD, denoted by \(\beta(\text{iRMSD})\). Thus, during training, both the forward flexing process and the score model are conditioned on \(\beta(\text{iRMSD})\) via the input \(\alpha(t;\beta(\text{iRMSD}))\). In the sampling phase, a simple EGNN model predicts the iRMSD from the two unbound structures, allowing the flexing schedule to adapt to the required level of conformational change for each sample.\\

Thus, our approach provides an efficient and adaptive mechanism to guide residue-level transformations toward physically plausible docking conformations.

\subsubsection{Score and Confidence Model}

The aforementioned global and local diffusion processes are coupled through a common, score and confidence model that preserves translational and rotational invariance or equivariance for various outputs (see summary in Table~\ref{table:equivariance_properties}).    

\begin{table}[h]
\centering
\caption{Transformation properties of score model outputs.}
\begin{tabular}{lcc}
\toprule
\textbf{Outputs} & \textbf{Translation} & \textbf{Rotation} \\
\midrule
$R$, $\mathbf{t}$ & Invariant & Equivariant \\
$R_{\text{rec}}$, $\mathbf{t}_{\text{rec}}$, $R_{\text{lig}}$, $\mathbf{t}_{\text{lig}}$ & Invariant & Equivariant \\
clddt & Invariant & Invariant \\
\bottomrule
\end{tabular}
\label{table:equivariance_properties}
\end{table}

\textbf{Intra- and Inter-Protein Graph Representation}.  We represent a protein complex structure as two types of graphs --- intra-protein (receptor or ligand) and inter-protein (receptor to ligand or vice versa) --- with nodes corresponding to individual residues and edges corresponding to spatial proximity.  

By default, we include protein features such as sequence (one-hot encoding for node features), evolution (residue-level ESM-2 embedding of protein sequence for node features), and structure ($C_\alpha$ coordinates and residue-frame orientations for node features and radial basis functions of residue--residue distances for edge features).  Due to the space limit, we include details for graph representations in the Appendix~\ref{sec:appendixB}.  

\textbf{Additional Features of Protein Dynamics}. 
% In graph representations of protein complex structures, we additionally include protein dynamics from normal mode analysis (NMA) as node and edge features for intra-protein graphs.  
To effectively model intra-protein flexibility and enhance the predictive accuracy of our adaptive residual flow matching approach, we incorporate residue-level flexibility features derived from Normal Mode Analysis (NMA). NMA provides intrinsic insights into protein flexibility by characterizing collective motions through normal modes, offering biologically meaningful priors for residue-level local flexibility. We provide a preliminary in the Appendix~\ref{sec:prelim-nma}.  

\begin{enumerate}
    \item Mean Square Fluctuation as Node Features.  

The Mean Square Fluctuation (MSF) quantifies the intrinsic flexibility of a residue by measuring its displacement across a set of normal modes. Specifically, for residue \( i \), the MSF is computed as:
\[
\text{MSF}_i = \sum_{m=1}^{M}\frac{1}{\lambda_m}\left|\mathbf{u}_i^m\right|^2,
\]
where \( \lambda_m \) denotes the eigenvalue corresponding to the \(m\)-th normal mode (eigenvector), and \(\mathbf{u}_i^m\) represents the displacement vector of residue \( i \) in the \(m\)-th mode. This scalar value serves as a node-level feature, providing biologically interpretable guidance to the flexibility-aware docking model.

\item Cross-Correlation of Residue Displacements as Edge Features 

Residue movements are inherently cooperative rather than independent. To capture such correlated fluctuations, we define residue-level cross-correlation values based on their displacement vectors from normal mode analysis (NMA). Specifically, for residues \( i \) and \( j \), the cross-correlation \( C(i,j) \) is defined as:
\[
C(i,j) = \frac{\sum_{m=1}^{M}\frac{1}{\lambda_m}\left(\mathbf{u}_i^m\cdot\mathbf{u}_j^m\right)}
{\sqrt{\sum_{m=1}^{M}\frac{1}{\lambda_m}\left|\mathbf{u}_i^m\right|^2 \cdot \sum_{m=1}^{M}\frac{1}{\lambda_m}\left|\mathbf{u}_j^m\right|^2}},
\]
where \(\lambda_m\) denotes the eigenvalue of the \( m \)-th normal mode. The resulting cross-correlation matrix \(C(i,j)\) serves as an edge-level feature, effectively encoding correlated residue motions critical to docking processes.

\end{enumerate}

Integrating these normal mode-derived features—residue-level MSF as node attributes and displacement cross-correlations as edge attributes—enables our docking model to explicitly incorporate inherent protein flexibility. This facilitates the learning of physically realistic and coordinated residue-level transformations, leading to improved accuracy and biological relevance of predicted docking conformations.

\begin{figure}[htbp]
    \centering
    \includegraphics[width=\linewidth]{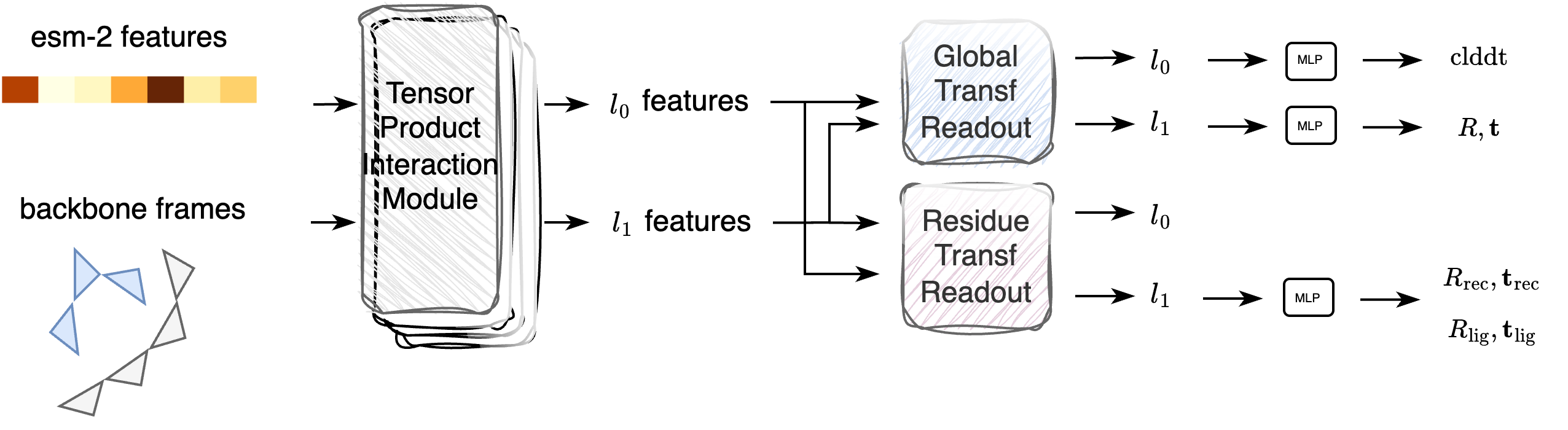}
    \caption{\textbf{Schematic representation of the score model architecture.} The interaction module, global transformation readout, and residue-level transformation readout are each composed of tensor product layers to ensure rotational equivariance and translation invariance.}
    \label{fig:arc}
\end{figure}

\textbf{Score Model Architecture}.  We use a common score model that is given the input of aforementioned intra- and inter-protein graph representations and outputs global and local translational and rotational scores as well as confidence prediction (quality estimation in clddt).  It consists of three modules of embedding, (tensor product) interaction, and output (Figure~\ref{fig:arc}).   

For the embedding module, we concatenate aforementioned node features with sinusoidal encoding of diffusion time $\tau$ and embed the node features with MLPs.  We similarly embed intra and inter-protein edge features with two MLPs.  For the interaction module, we define inter-residue messages as tensor products of current node features and spherical harmonic representations of the edge vectors (up to degree $\ell=2$); and we update node features layer by layer by pooling such messages. For the output module, we limit the spherical harmonic edge vectors to degree $\ell=0$ or $\ell=1$, before feeding them to corresponding MLPs, depending on whether the output is invariant or equivariant to rotations of the initial protein complex structure at $\tau=1$.   
% based SE(3)-Equivariant score model $s_\theta((\mathbf{x},O), \tau)$ taking in the backbone frame of each residue in the receptor and ligand proteins at time $\tau$. In the score model, the two protein structures (receptor protein and ligand protein) are further represented in geometric graphs formed by protein residues (each associated with its central carbon coordinates). 
The model architecture is illustrated in Figure~\ref{fig:arc} and model outputs equivariance properties are showed in Table ~\ref{table:equivariance_properties}. Due to the space limit, we include details for the score model in the Appendix~\ref{sec:appendixB}.\\

\textbf{Confidence score} We use 0.5\AA, 2\AA, 4\AA $ $ and 8\AA $ $ for the clddt cutoffs. The contact lddt(clddt) prediction (as part of the invariant output of the score model) is used for confidence prediction to select samples from the results of the sampling process. We employ out-of-distribution input pairs of the form \((\left
(x, O)_\epsilon^{\mathrm{pred}}, 1 - \epsilon\right), \text{with } \epsilon = \frac{1}{\# \text{sampling step}}, \)
as inputs to the score model \( s_\theta((x, O), \tau) \). This design choice is motivated by the fact that for the clddt prediction, the high-dimensional structural information in \( (x, O) \) predominates, while the low-dimensional time parameter \( \tau \) is deliberately set near an extreme value. Consequently, the influence of \( \tau \) is minimized, effectively isolating the contribution of the predicted structure \( (x, O) \) to the score. We validated this approach using the DB5.5 validation set, which yielded a moderate average-cross-sample Pearson correlation coefficient of \(-0.286\) between the iRMSD and the predicted clddt score over 40 sampling runs.

\subsubsection{Training Loss}

\[
\mathcal{L}=w_{\text{diff}}\mathcal{L}_{\text{diff}} + w_{\text{iFAPE}}\mathcal{L}_{\text{iFAPE}}+w_{\text{lddt}}\mathcal{L}_{\text{lddt}}
\]

\(\mathcal{L}_{\text{diff}}\) is the weighted sum of every sub-diffusion loss including the ligand global rotation and translation variance-weighted score-matching losses, residue-level rotations and translations flow matching losses for receptor and ligand separately (see details in the Appendix~\ref{sec:appendixC}). \(\mathcal{L}_{\text{iFAPE}} \) is the interface focused FAPE loss which is in the distance sense. \(\mathcal{L}_{\text{lddt}}\) is the contact lddt l2 loss. Details for training and hyperparameter tuning are in the Appendix~\ref{sec:appendixD}.  \\

\subsection{Data}

\subsubsection{DIPS-AF for Pre-training}
We curated our dataset DIPS-AF from two datasets: AlphaFold Protein Structure Database (AlphaFold DB) [version 2022-11-01] \cite{varadi2022alphafold} to simulate the unbound structures of individual proteins and the Database of Interacting Protein Structures (DIPS) \cite{townshend2019end} to retrieve diverse and representative bound structures of paired proteins. AlphaFold DB provides extensive and high-accuracy protein-structure predictions with UniProt \cite{uniprot2015uniprot} ID as the reference. DIPS consists of 42,826 binary protein complexes dataset referenced with PDB \cite{burley2017protein} id and chain ids for both of the chains involved in each of the complex.  \\

For each chain in the complex, resulted from matching its associated PDB id and chain id to the UniProt id with PDB GraphQL-based API \cite{rose2021rcsb} by using query with entry\_ids (PDB id) for polymer\_entities information including a matching UniProt accession for each chain and then matching the chain to the AlphaFold DB referenced with UniProt id, 39,251 of complexes have predicted structures for both chains in the AlphaFold DB. We labeled each protein pair in the DIPS as bound receptor and bound ligand structures and further augmented the pair with two coresponding matched structures in AlphaFold DB as the unbound structures. Now we have four structures for each complex namely unbound and bound receptors ($ur, br$), unbound and bound ligands ($ul, bl$). \\

As the sequences between unbound or bound protein structures could be  non-identical, we further processed the two pairs of unbound-bound structures by  local sequence alignment with BLOSUM62 substitution matrix \cite{eddy2004did} for each pair and kept the locally aligned segments of sequences and their corresponding structures. Since the kept aligned sequence might be only parts of the receptor/ligand sequences, we only kept the cases whose average percentage of aligned sequence length in the bound sequence length among receptor and ligand is  at least $80\%$ to ensure the quality of the data.  This results in nearly 39,000 cases with various docking difficulty levels (statistics shown in Table~\ref{tab:dips_af_stats}). \\

\begin{table}[h]
\caption{\textbf{Statistics of the DIPS-AF dataset categorized by docking difficulty.} The classification of docking difficulty is following the same criteria of conformational change levels as in the DB5.5 dataset.}
\centering
\begin{tabular}{l r}
\hline
\textbf{Category} & \textbf{Number of Complexes} \\
\hline
Rigid-body   & 30,723 \\
Medium       & 3,320 \\
Difficult    & 4,844 \\
\hline
\textbf{Total} & \textbf{38,887} \\
\hline
\end{tabular}

\label{tab:dips_af_stats}
\end{table}

\subsubsection{DB5.5 for Finetuning and Testing}

We adopted commonly-used DB5.5~\cite{vreven2015updates} to benchmark docking performances, including its data splits of training, validation, and test sets~\cite{diffdock_pp_ref}.  In particular, DB5.5's test set contains 25 protein--protein pairs, including 16 rigid, 6 medium, and 3 difficult cases that correspond to increasing levels of conformational changes and docking difficulty.  

There is no sequence homology between DIPS-AF and DB5.5 datasets due to the way DIPS was constructed.  Specifically, DIPS excludes any complex that has any individual protein with over 30\% sequence identity when aligned to any protein in DB5 to eliminate the leakage between the two datasets. \cite{townshend2019end}

% \section{Result and Discussion}
% We evaluated the performance of our proposed hierarchical adaptive flexible docking approach on the DB5.5 test set using the DIPS-AF training set for model development. 

% \begin{table}[htbp]
% \caption{\textbf{Table for model index.} Every model comparing to the above one is incremental; Only one aspect is added or changed to have comparison. }
% \centering
% \begin{tabular}{lc}
% \toprule
% \textbf{Experiments Description} & \textbf{Model} \\
% \midrule
% DIPS-AF ($\beta=5$) & Model0 \\
% DIPS-AF ($\beta=5$); NM features & Model\_NMFeature \\
% DIPS-AF (adaptive $\beta$); NM features & Model\_adaptSchedule \\
% DB5 (adaptive $\beta$); NM features & Model\_adaptSchedule\_noPretrain \\
% \bottomrule
% \end{tabular}

% \label{tab:model_index}
% \end{table}

\section{Results and Discussion}
We evaluate the performance of our hierarchical adaptive flexible docking approach on the DB5.5 test set, using the DIPS-AF training set for model development. We compare our model against DiffDock-PP and GeoDock. Additionally, we conduct ablation studies to assess the contribution of key components such as Normal Mode-derived (NM) features and our adaptive flexibility schedule.

\subsection{Model Performance Comparison}
\subsubsection{Overall Comparison}
\begin{figure}[htbp]
    \centering
    \includegraphics[width=0.8\linewidth]{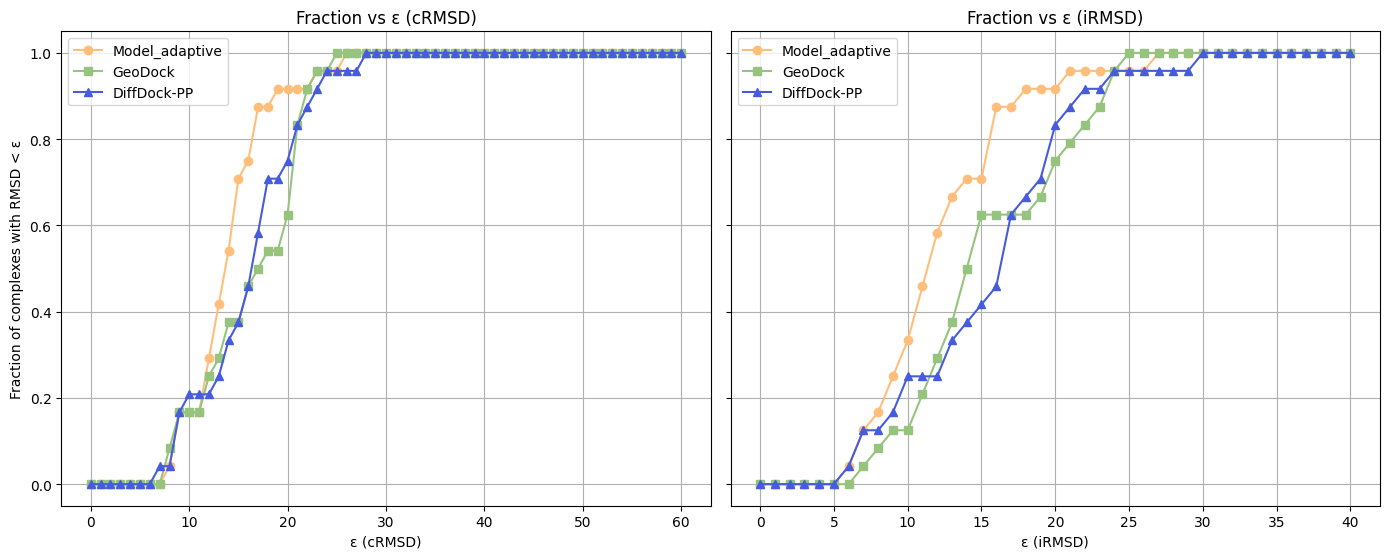}
    \caption{Fraction of samples with RMSD < $\epsilon$ for different $\epsilon$ values (DB5.5 Test Set, 24 Samples, Excluding 1N2C) }
    \label{fig:fraction_less_epsilon}
\end{figure}

Table~\ref{tab:db5_models_comparison} summarizes the Complex RMSD (cRMSD) and Interface RMSD (iRMSD) results across different models. Our model (Model\_adaptive) consistently outperforms DiffDock-PP and GeoDock, achieving lower mean and median values for both cRMSD and iRMSD. This conclusion can also be drawn in the cumulative distributions shown in  Figure~\ref{fig:fraction_less_epsilon} (the fraction of samples with RMSD < $\epsilon$ for different $\epsilon$ values).   Our model's distributions are mostly above the two other models' for both cRMSD and iRMSD metrics.

In terms of inference speed, our flexible diffusion model is as fast the  rigid-body diffusion model DiffDock-PP and is 2.6-times slower compared to the non-diffusion flexible docking model GeoDock (which uses an iterative transformer archicture similar to AlphaFold 2).  
% The success rate metrics in Figure~\ref{fig:modelAda_vs_GeoDock_diffdock_dockingHit} further indicate that our approach is more effective than the other two methods with full rigidity and full flexibility assumptions respectively.

\begin{table}[htbp]
    \setlength{\tabcolsep}{3pt} % Adjust column spacing
    \caption{\textbf{DB5.5 Test Set Evaluation Top-1 Results: Model Comparison.}}
    \label{tab:db5_models_comparison}
    \begin{tabular}{lccc|ccc|c}
    \toprule
    & \multicolumn{3}{c}{Complex RMSD (Å)} & \multicolumn{3}{c}{Interface RMSD (Å)} & \multicolumn{1}{c}{Runtime (s)}\\
    \cmidrule(lr){2-4} \cmidrule(lr){5-7} \cmidrule(lr){8-8}
    \textbf{Methods} & Mean$\pm$Std & Median & \%($<$10)  & Mean$\pm$Std & Median & \%($<$10) & Mean \\
    \midrule
        GeoDock               & 15.51$\pm$4.5  & 15.62 & 16  & 14.84$\pm$5.1  & 13.83 & 12 & 2.24 \\
    DiffDock-PP         & 16.28$\pm$5.3  & 16.52 & 20  & 16.04$\pm$6.4  & 16.23 & 24 & 5.92 \\
    Model\_adaptive      & \textbf{13.98}$\pm$4.4 & \textbf{13.88} & 20  & \textbf{12.88}$\pm$4.6 & \textbf{11.93} & 28 & 5.52 \\
    \bottomrule
    \end{tabular}
\end{table}

\subsubsection{Comparison Based on Docking Difficulty (Conformational Change Levels)}

% \YS{I should emphasize Figure 3 and cite details in Table 7 for the difficulty-level split performances. Possibly move it in and remove \%<5 in Table 3.} 

\begin{figure}[htbp]
    \centering
    \includegraphics[width=0.7\linewidth]{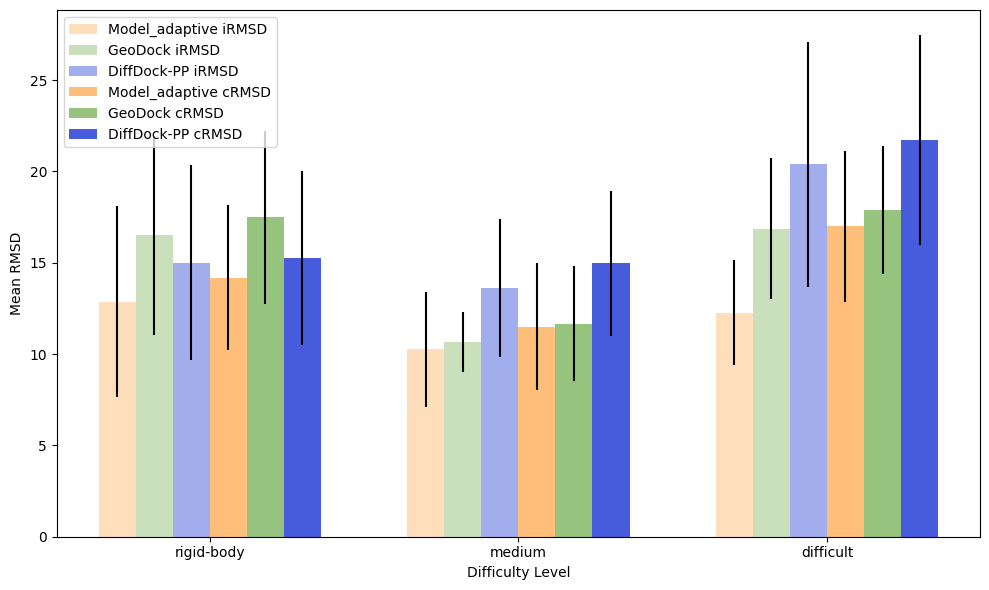}
    \caption{\textbf{DB5.5 Test Set Evaluation Top-1 Results Figure By Difficulty Level: Model Comparison.} DB5.5 test set mean complex RMSD and mean Interface RMSD results comparison between ours (model\_adaptSchedule), GeoDock and DiffDock-PP for samples on different difficulty levels. We exclude 1N2C evaluation results in the medium level, since GeoDock failed to generate 1N2C prediction.}
    \label{fig:modelAda_vs_GeoDock_diffdock_diffLevels}
\end{figure}

% \YS{Table 7 in Appendix D move ahead? }
Table~\ref{tab:db5_model_comparison_difficulty_level} and Figure~\ref{fig:modelAda_vs_GeoDock_diffdock_diffLevels} show that our model outperforms both GeoDock and DiffDock-PP across all difficulty levels, achieving the lowest mean values for both complex and interface RMSD. 
compare model.  This result confirms that our method adapts well to varying degrees of conformational changes. This is attributed to our adaptive flexing schedule, which allows the model to introduce flexibility adaptive to different cases based on the predicted interface RMSD. Consequently, our approach surpasses rigid docking methods in cases requiring minimal conformational change while also outperforming fully flexible models in highly flexible cases—demonstrating its ability to generalize across docking scenarios that were previously addressed by separate, specialized models.

\begin{table}[htbp]
    \setlength{\tabcolsep}{3pt} % Adjust column spacing
    \caption{\textbf{DB5.5 Test Set Evaluation Top-1 Results Table By Difficulty Level: Model Comparison.} Bold numbers indicate the lowest (best) mean RMSD values, while underlined  numbers represent the second lowest values in each column. *For medium level group, GeoDock is missing the 1N2C sample due to out of memory issue when evaluation; 5 cases (out of 6) are considered instead.}
    \label{tab:db5_model_comparison_difficulty_level}
    \begin{tabular}{lccc|ccc}
    \toprule
    & \multicolumn{3}{c}{Complex RMSD Mean$\pm$Std (Å)} & \multicolumn{3}{c}{Interface RMSD Mean$\pm$Std (Å)} \\
    \cmidrule(lr){2-4} \cmidrule(lr){5-7} 
     & rigid  & medium  & difficult  & rigid  & medium  & difficult  \\
    \textbf{Methods} & (16 cases)  & (5 cases*)  & (3 cases)  & (16 cases)  & (5 cases*)  & (3 cases)   \\
    \midrule
    GeoDock             & 17.49$\pm$4.74  & \underline{11.65$\pm$3.15}   & \underline{17.90$\pm$3.52}   & 16.51$\pm$5.45  & \underline{10.65$\pm$1.66}   & \underline{16.87$\pm$3.86}   \\
    DiffDock-PP        & \underline{15.25$\pm$4.75}   & 14.97$\pm$3.98  & 21.72$\pm$5.75  & \underline{14.99$\pm$5.34}   & 13.62$\pm$3.79  & 20.40$\pm$6.71  \\
        Model\_adaptive      & \textbf{14.18$\pm$3.98}  & \textbf{11.50$\pm$3.49}  & \textbf{16.99$\pm$4.14}  & \textbf{12.86$\pm$5.22}  & \textbf{10.26$\pm$3.15}  & \textbf{12.26$\pm$2.87} \\
    \bottomrule
    \end{tabular}
\end{table}

% In detail, our model outperforms both GeoDock and DiffDock-PP across all difficulty levels and both metrics, achieving the lowest mean RMSD values for both complex and interface RMSD. This result further highlights the effectiveness of the adaptive training strategy, which is designed to optimize the model's ability to generalize and handle various docking scenarios. 

In comparison, GeoDock, which is specifically designed for flexible protein-protein docking, shows strong performance in medium and difficult cases but relatively weak performance in rigid cases. Its design allows it to handle flexibility, which is crucial when proteins undergo significant conformational changes during docking. However, while GeoDock performs competitively in medium and difficult scenarios, it is slightly outperformed by our model across all difficulty levels. This suggests that while GeoDock is tailored for flexible cases, it does not generalize as effectively in more rigid tasks, where our model performs better.

DiffDock-PP, as a model optimized for rigid-body protein docking, excels in rigid-body cases but faces limitations in handling flexible docking scenarios. For medium and difficult cases, which often require flexibility handling, DiffDock-PP's performance drops significantly, especially in the difficult category. This highlights that DiffDock-PP is not suitable for more complex docking tasks involving flexibility.

Therefore, while GeoDock excels at handling flexible docking and DiffDock-PP is effective for rigid-body docking, our model offers a more versatile and robust solution. The adaptive strategy enables it to perform better across a broader range of scenarios, making it a more effective model for a variety of docking challenges.

\subsubsection{Case-by-Case Comparison and Statistical Significance}

\begin{figure}[htbp]
    \centering
    \includegraphics[width=\linewidth]{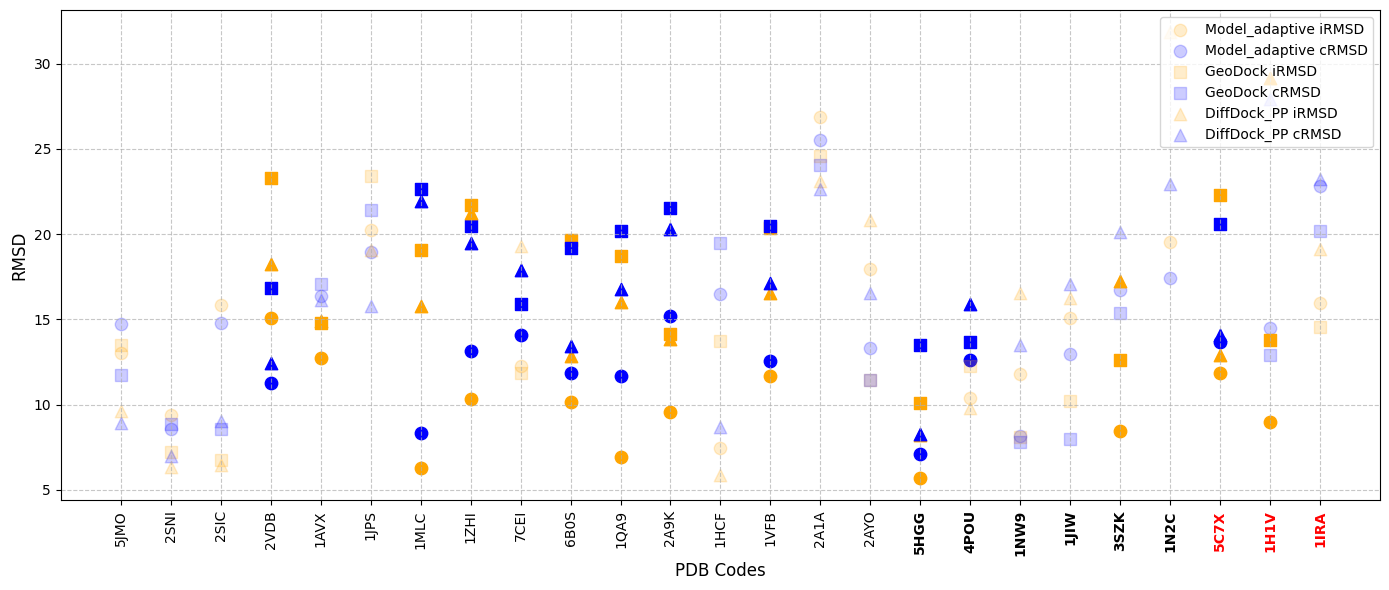}
    \caption{DB5.5 test set case-by-case complex RMSD and Interface RMSD results comparison among ours (model\_adaptive), GeoDock and DiffDock-PP based on top 1 results. The normal font PDBs are for rigid cases and the bold PDBs are for medium cases while the red font PDBs are for difficult cases. For each case, either regarding cRMSD or iRMSD, if our model is better than both the other two models, we use more opaque colors for those dots.}
    \label{fig:modelAda_vs_GeoDock_diffdock_cbc}
\end{figure}

Additionally, Figure~\ref{fig:modelAda_vs_GeoDock_diffdock_cbc} presents a case-by-case performance analysis; Our model is best for 48\%, second best for 40\% in cRMSD and is best for 52\%, second best for 36\% in iRMSD. Our model also significantly outperforms both GeoDock (cRMSD: p = 0.0282, iRMSD: p = 0.0157) and DiffDock‐PP (cRMSD: p = 0.0258, iRMSD: p = 0.0044) in a paired one‐sided Wilcoxon signed‐rank test.

\subsection{Ablation Study}

To understand the contribution of each component in our model, we conduct an ablation study using four configurations:

\begin{itemize}
    \item \textbf{Model0.} Our base model using hierarchical transformations.
    \item \textbf{Model\_NMFeatures.} Adding Normal Mode (NM) features to enhance residue flexibility representation.
    \item \textbf{Model\_adaptive.} Incorporating an adaptive flexibility schedule.
    \item \textbf{Model\_adaptive\_noPretrain.} Training only on DB5.5, without pretraining on DIPS-AF.
\end{itemize}

The results in Table~\ref{tab:db5_ablation_study} demonstrate the effectiveness of our hierarchical formulation, as Model0 already outperforms both GeoDock and DiffDock-PP. Adding NM features further improves accuracy, while adaptive scheduling provides the most significant performance boost by dynamically adjusting the flexibility level for each protein complex.

Interestingly, Model\_adaptive\_noPretrain performs significantly worse than Model\_adaptive, highlighting the benefits of pretraining on DIPS-AF, despite the high heterogeneity between AlphaFold-predicted unbound structures and DB5.5 experimental unbound structures.

% \begin{table}[htbp]
%     \setlength{\tabcolsep}{3pt} % Adjust column spacing
%     \caption{\textbf{DB5.5 Test Set Evaluation Top-1 Results: Ablation Study.}}
%     \label{tab:db5_ablation_study}
%     \begin{tabular}{lcccc|cccc}
%     \toprule
%     & \multicolumn{4}{c}{Complex RMSD (Å)} & \multicolumn{4}{c}{Interface RMSD (Å)} \\
%     \cmidrule(lr){2-5} \cmidrule(lr){6-9} 
%     \textbf{Methods} & Mean$\pm$Std & Median & \%($<$5) & \%($<$10)  & Mean$\pm$Std & Median & \%($<$5) & \%($<$10)  \\
%     \midrule
%     Model0              & 15.59$\pm$5.3  & 14.45 & 4 & 20  & 13.78$\pm$4.7  & 13.39 & 4 & 24  \\
%     Model\_NMFeatures    & 15.42$\pm$5.6  & \textbf{13.68} & 4 & 16  & 13.15$\pm$4.6  & 12.58 & 4 & 24  \\
%     Model\_adaptive      & \textbf{13.98}$\pm$4.4 & 13.88 & 0 & 20  & \textbf{12.88}$\pm$4.6 & \textbf{11.93} & 0 & 28  \\
%     (Model\_adaptive\_noPretrain) & 16.62$\pm$4.5 & 15.82 & 0 & 12  & 14.42$\pm$3.7  & 14.09 & 0 & 12  \\
%     \bottomrule
%     \end{tabular}
% \end{table}

\begin{table}[htbp]
    \setlength{\tabcolsep}{4pt} % Adjust column spacing
    \caption{\textbf{DB5.5 Test Set Evaluation Top-1 Results: Ablation Study.}}
    \label{tab:db5_ablation_study}
    \begin{tabular}{lccc|ccc}
    \toprule
    & \multicolumn{3}{c}{Complex RMSD (Å)} & \multicolumn{3}{c}{Interface RMSD (Å)} \\
    \cmidrule(lr){2-4} \cmidrule(lr){5-7} 
    \textbf{Methods} & Mean$\pm$Std & Median & \%($<$10)  & Mean$\pm$Std & Median & \%($<$10)  \\
    \midrule
    Model0              & 15.59$\pm$5.3  & 14.45 & 20  & 13.78$\pm$4.7  & 13.39 & 24  \\
    Model\_NMFeatures    & 15.42$\pm$5.6  & \textbf{13.68} & 16  & 13.15$\pm$4.6  & 12.58 & 24  \\
    Model\_adaptive      & \textbf{13.98}$\pm$4.4 & 13.88 & 20  & \textbf{12.88}$\pm$4.6 & \textbf{11.93} & 28  \\
    (Model\_adaptive\_noPretrain) & 16.62$\pm$4.5 & 15.82 & 12  & 14.42$\pm$3.7  & 14.09 & 12  \\
    \bottomrule
    \end{tabular}
\end{table}

\subsection{Remaining Gaps in Scoring and Sampling Docking Conformations}
To reveal performance gaps, we first assess confidence estimationby comparing our predicted clddt scores against the true iRMSD values.

\begin{figure}[htbp]
    \centering
    \includegraphics[width=0.5\linewidth]{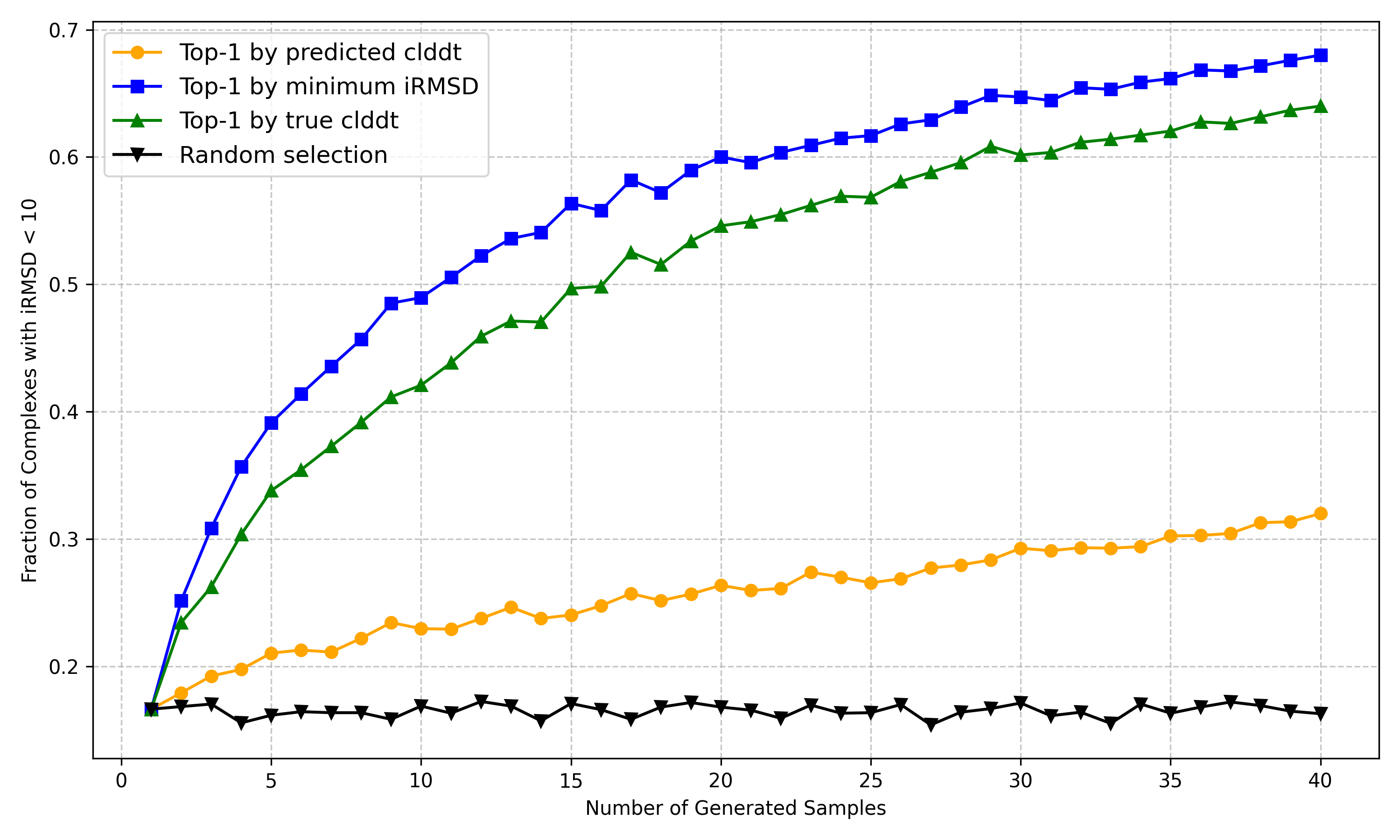}
    \caption{Top-1 Performance vs. Number of Generated Samples for the DB5.5 Test Set. The plot shows the fraction of complexes achieving an iRMSD below 10 \AA$ $ as a function of the number of generated samples (from 1 to 40), averaged over 100 random trials. The orange curve represents the performance when selecting the candidate with the highest \textbf{predicted clddt score}, and the green curve represents the performance when selecting the candidate with the highest \textbf{true clddt score} while the blue curve corresponds to choosing the candidate with the lowest \textbf{iRMSD} and the black curve is to choose the candidate \textbf{randomly}.}
    \label{fig:plddt_eval_top1_by_tlddt_plddt_vs_by_best_iRMSD}
\end{figure}

As shown in Figure~\ref{fig:plddt_eval_top1_by_tlddt_plddt_vs_by_best_iRMSD}, our current confidence predictor leaves much room for improvement. While our predicted clddt ranks plausible docking poses better than random selection, there remains a gap between the best-selected (orange) and best-possible predictions (green). This suggests that further finetuning on generated samples by the sampling process (rather than only trained on the samples in the forward process) could enhance ranking performance. Moreover, the difference between best-possible predictions measured by clddt (green) versus iRMSD (blue) indicates that introducing a confidence metric prediction directly tied to iRMSD could be beneficial.

\begin{figure}[htbp]
    \centering
    \includegraphics[width=0.6\linewidth]{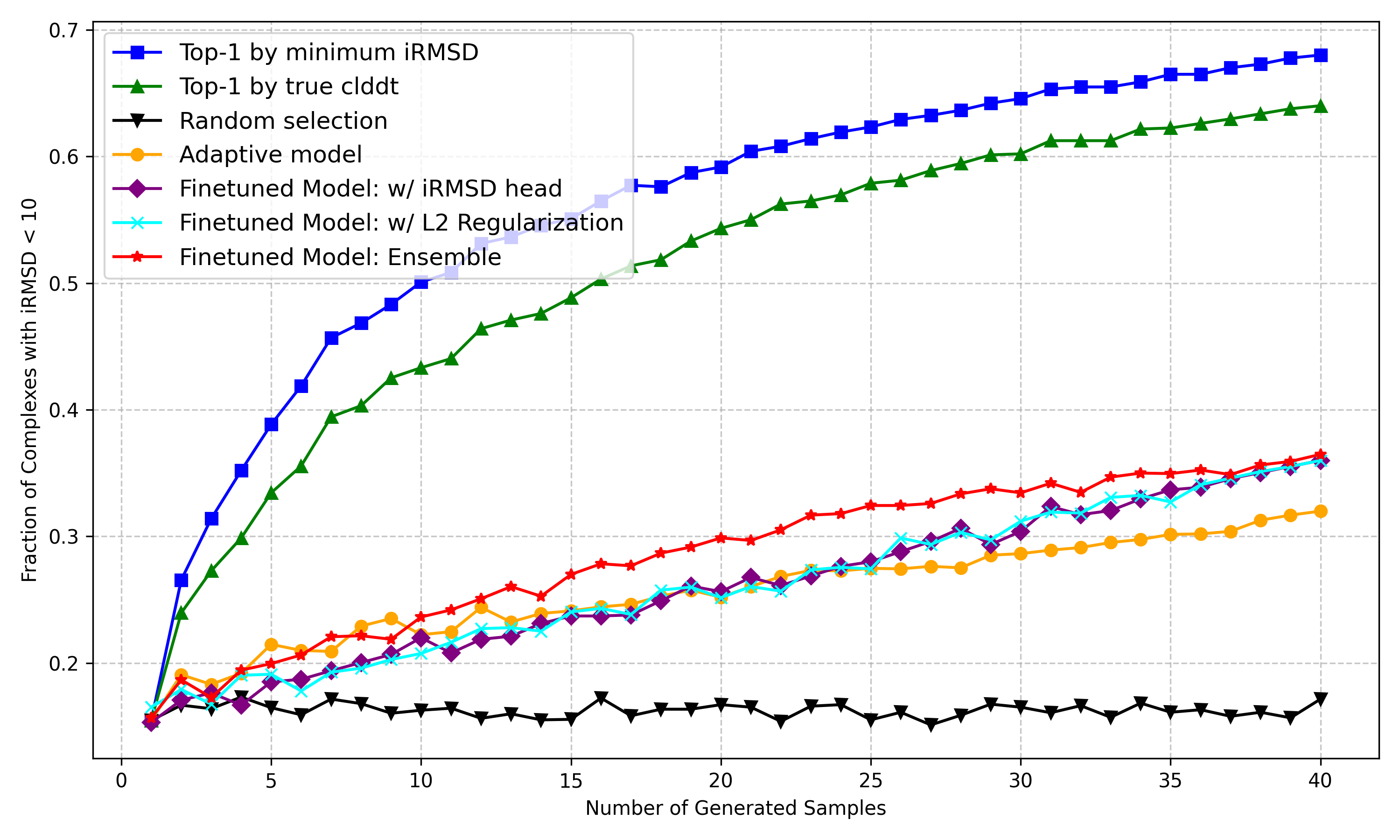}
    \caption{Top-1 Performance vs. Number of Generated Samples for the DB5.5 Test Set. The plot shows the fraction of complexes achieving an iRMSD below 10 \AA{} as a function of the number of generated samples (from 1 to 40). The blue curve represents selecting the candidate with the \textbf{minimum iRMSD} (ideal selection), while the green curve shows selection by \textbf{true clddt score}. The black curve indicates \textbf{random selection}. The orange curve corresponds to the \textbf{adaptive model} prediction, the purple curve to the \textbf{finetuned model with an additional iRMSD prediction head}, the cyan curve to the \textbf{finetuned model with L2 regularization}, and the red curve to the \textbf{ensemble model of finetuned models}.}
    \label{fig:finetune_overall_results}
\end{figure}

To explore this idea, we finetune two additional confidence models on DB5.5 samples generated by the trained adaptive\_model, along with their ensemble. The two finetuned models include: (1) a model with an additional binary prediction head for iRMSD < 10 \AA, and (2) a model trained with a dropout rate of 0.3 and L2 regularization to optimize generalization. As shown in Figure~\ref{fig:finetune_overall_results}, both finetuned models, as well as their ensemble, achieve slightly better performance than the original adaptive\_model, providing initial validation for our hypothesis. Further validation by finetuning on the full combined dataset (DIPS-AF and DB5.5 samples) should help to close the performance gap further. Moreover, during finetuning, we observe clear overfitting, suggesting that designing a more tailored and simpler model, or incorporating more informative input features (such as side-chain modeling), could be beneficial. We leave these directions for future work.

We also noted that the absolute performances of the best-possible predictions measured by clddt (green) or iRMSD (blue) are not satisfactory either, which indicates that there is also much room to improve  diffusion-based sampling besides confidence-based scoring.

\subsection{Case Study}

We chose two docking cases, one rigid (PDB ID: 2SNI) and one medium (PDB ID: 5HGG), to investigate in-depth the dynamics of our learned diffusion models over global and local motions.  

\begin{figure}[htbp]
    \centering
    % First figure (on top)
    \begin{subfigure}{\linewidth}
        \centering
        \includegraphics[width=\linewidth]{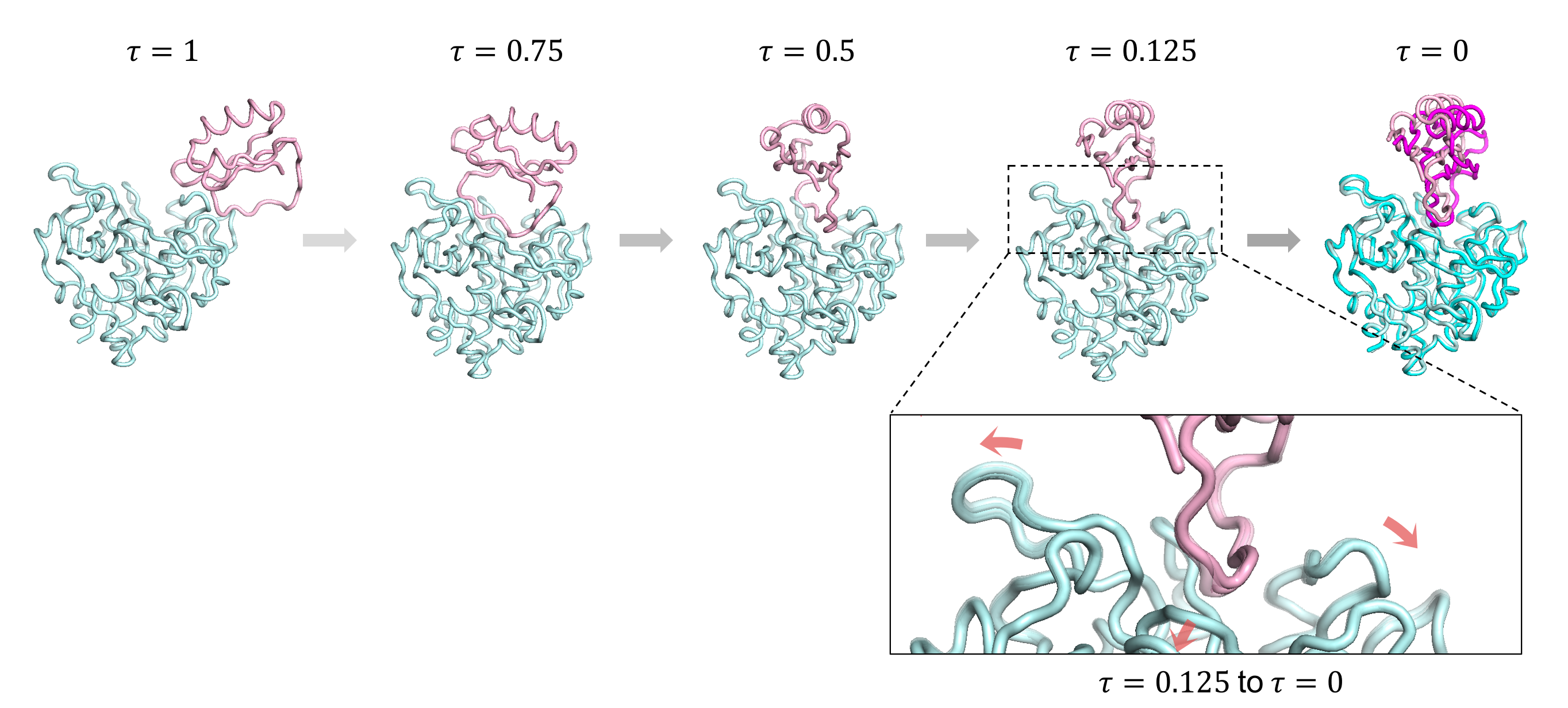}
        \caption{Docking process for case 2SNI.}
        \label{fig:top}
    \end{subfigure}
    \vspace{1em} % Optional vertical spacing between the two subfigures
    % Second figure (below)
    \begin{subfigure}{\linewidth}
        \centering
        \includegraphics[width=\linewidth]{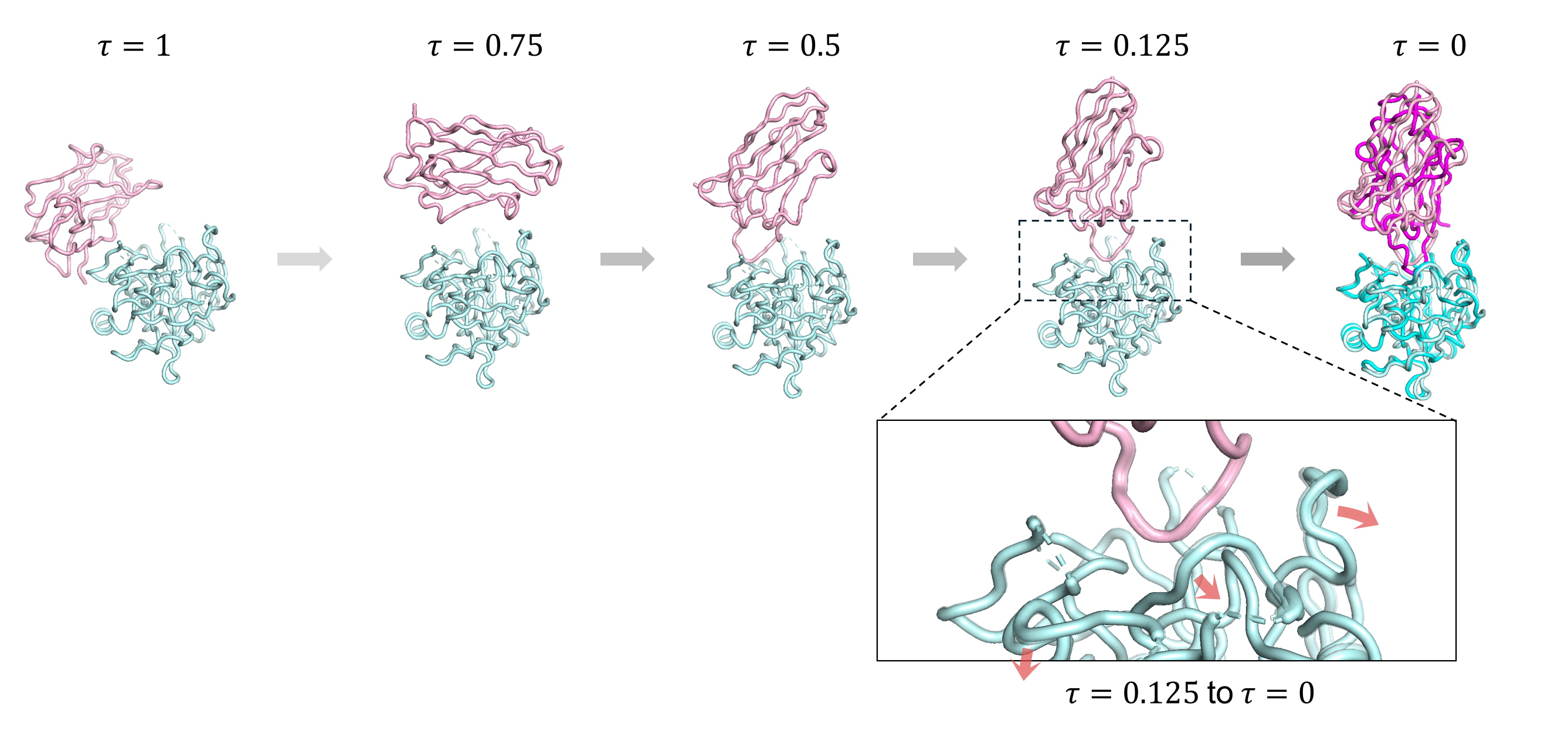}
        \caption{Docking process for case 5HGG.}
        \label{fig:bottom}
    \end{subfigure}
    \caption{The illustration depicts the docking process over time. Structures with light pink and pale cyan are for the predicted and ones with magenta and cyan are for the ground truth. In the zoomed-in image, cartoon with increased opacity indicates later stages. }
    \label{fig:case_study}
\end{figure}

As shown in Figure~\ref{fig:case_study},  for both cases the sampling process begins from the unbound state with randomly initialized positions and orientations. It then gradually refines the structure through two levels of motions: global (rigid-body) and local (residue-level flexibility). Notably, the magnitude of global motion decreases over time, while the extent of residue-level deformation increases. This behavior is intentionally designed to mimic the induced fit mechanism observed in protein docking, where local conformational changes occur as binding progresses.

Moreover, the predicted iRMSD governs the timing of residue-level deformations, enabling earlier local flexing in cases where significant conformational changes are anticipated. This adaptive mechanism allows the model to effectively recognize and shape protein binding pockets, facilitating accurate ligand engagement. However, in both cases shown—2SNI and 5HGG—some degree of mismatch remains: a positional misalignment in 2SNI and an orientation error in 5HGG.

These discrepancies may arise from multiple sources. A key limitation is the lack of explicit side-chain modeling in our current framework. Since contact interfaces are largely defined by side-chain interactions, omitting this information can hinder precise alignment, especially in tight or highly specific binding regions.

Another potential factor is the one-shot regression of the iRMSD used to determine the adaptive schedule. While effective in many cases, this estimation may introduce inaccuracies in more complex or ambiguous docking scenarios, leading to suboptimal timing for local flexibility activation.

Addressing these limitations is an important direction for future work. One direction is the incorporation of explicit side-chain modeling to improve interface accuracy, particularly in cases where precise steric and chemical complementarity is critical. In addition, instead of relying solely on a one-shot regression of the predicted iRMSD to determine the flexing schedule, future approaches could explore reinforcement learning (RL)-based strategies to dynamically adjust the schedule during sampling by modulating when and how much local flexibility is introduced.

\section{Conclusion}
We have presented a hierarchical adaptive flexible docking framework that integrates global rigid-body transformations and  fine-grained residue-level flexibility with an adaptive flexing schedule. Our approach models the protein complex on a homogeneous space of products of global and local $SE(3)$,
thereby ensuring consistent treatment of both global and local conformational variables. By incorporating normal mode-derived dynamics features (besides sequence, evolution and structure features) and adaptively adjusting the level of flexibility based on predicted interface RMSD, our method bridges the gap between rigid-body and fully flexible docking models.

Experimental results on the DB5.5 benchmark demonstrate that our model consistently outperforms state-of-the-art rigid and flexible diffusion  models in terms of complex RMSD, interface RMSD, and success rates. Ablation studies further confirm the significant contributions of normal-mode features, the adaptive scheduling strategy, and pretraing using the DIPS-AF dataset.  

In future work, we plan to refine the confidence estimation component and further enhance the model's ability to handle large-scale conformational changes, while extending the resolution from the backbone to all atoms,  including the side chain.

\appendix
\clearpage
% \section{Appendix}
\newpage
\section*{Appendix}
\addcontentsline{toc}{section}{Appendix}
\setcounter{section}{0}

\section{Preliminaries on Normal Mode Analysis}\label{sec:prelim-nma}

Normal Modes Analysis (NMA) is a powerful tool to study the intrinsic flexibility of macromolecules under the harmonic approximation. Starting from an equilibrium conformation \(\mathbf{x}_0\), the potential energy is locally approximated by
\[
V(\mathbf{x}) \approx V(\mathbf{x}_0) + \frac{1}{2}(\mathbf{x}-\mathbf{x}_0)^T H (\mathbf{x}-\mathbf{x}_0),
\]
where \(H = \nabla^2 V(\mathbf{x}_0)\) is the Hessian matrix. The normal modes are obtained by solving the eigenvalue problem
\[
H\,\mathbf{q}_i = \lambda_i\,\mathbf{q}_i,
\]
with eigenvalues \(\lambda_i\) representing the stiffness along the corresponding eigenvectors \(\mathbf{q}_i\). The system's displacement can then be expressed as a linear combination of these modes,
\[
\Delta \mathbf{x} = \sum_{i} a_i \,\mathbf{q}_i,
\]
and, the mean square fluctuation of a residue \(j\) is approximated by
\[
\text{MSF}_j \propto \sum_i \frac{(q_{ij})^2}{\lambda_i},
\]
where \(q_{ij}\) is the \(j\)th component of \(\mathbf{q}_i\). Typically, only the first few nontrivial modes (i.e., those with the smallest nonzero eigenvalues) are considered, as they capture the most significant, biologically relevant motions.

\section{Score Model Architecture} \label{sec:appendixB}
To ensure both global protein-level and local residue-level transformations outputs invariant to translation and equivariant to rotation of the whole complex, we employ the common center of mass subtraction strategy and utilize the following tensor-product neural networks to get the SO(3)-equivariant outputs.
Following the formulation of DiffDock, the architectures can be decomposed into three main parts: embedding layer, interaction layers, and output layer.

\textbf{Graph Representation} We define four graphs with each graph having notation of $\mathcal{G} = (\mathcal{V}, \mathcal{E})$. Each node $i \in \mathcal{V}$ in each of the four graphs $\mathcal{G}$ represents one residue and has 3D coordinates $\mathbf{x}_i$ for its central carbon $C\alpha$ position. Each edge in each of the four graphs is given by the neighbors defined with a designated $C\alpha$-$C\alpha$ Euclidean distance cutoff specific to the graph type.

\textbf{}The four graphs are categorized into two types: inter- graphs $\{\mathcal{G}_{rl}, \mathcal{G}_{lr}\}$ and intra- graphs $\{\mathcal{G}_{rr}, \mathcal{G}_{ll}\}$. 

\begin{itemize}
    \item For intra- graphs $\{\mathcal{G}_{rr} = (\mathcal{V}_{r}, \mathcal{E}_{rr}), \mathcal{G}_{ll} = (\mathcal{V}_{l}, \mathcal{E}_{ll})\}$, they are defined with nodes $\mathcal{V}_{r}$ or $\mathcal{V}_{l}$ as all the residues within the receptor protein or ligand protein and edges $\mathcal{E}_{rr}$ or $\mathcal{E}_{ll}$ as the edges within the radius graph for the set of nodes with a cutoff 10\AA.
        
    \item For inter- graphs $\{\mathcal{G}_{rl}= (\mathcal{V}_{rl}, \mathcal{E}_{rl}), \mathcal{G}_{lr} = (\mathcal{V}_{lr}, \mathcal{E}_{lr})\}$, they are defined with edges $\mathcal{E}_{rl} = \{e_{i \rightarrow j} | i \in \mathcal{V}_{r} \And   \forall j \in \mathcal{V}_{l} \; .s.t \; ||\mathbf{x}_i - \mathbf{x}_j|| < (40 + 3*\sigma_{tr}) \text{\AA} \}$ and $\mathcal{E}_{lr} = \{e_{i \rightarrow j} |i \in \mathcal{V}_{l} \And   \forall j \in \mathcal{V}_{r} \; .s.t \; ||\mathbf{x}_i - \mathbf{x}_j|| < (40 + 3*\sigma_{tr}) \text{\AA}\}$  and nodes $\mathcal{V}_{rl}$ or $\mathcal{V}_{lr}$ as any node constituting the edges $\mathcal{E}_{rl}$ or $\mathcal{E}_{lr}$.
\end{itemize}

\textbf{Features and Notation} For each node $i$ representing one residue of amino acid type $aa_{i}$ in each graph, there is $C\alpha$ coordinates $\mathbf{x}_i$ associated. We define for node features $\mathbf{h}_i$ and edge features $e_{ij}$:

\begin{itemize}
    \item For any node $i$, node features $\mathbf{h}_i$ is the concatenation of the scalar features $\mathbf{s}_i$ (invariant/type-$0$ features) and the vector features $V_i$ (equivariant/type-$\ell$ ($\ell>0$) features), i.e.\ $\mathbf{h}_i = (\mathbf{s}_i, V_i)$.
    \item For any edge $e_{i \rightarrow j}$, edge features is represented as $e_{ij}$.
\end{itemize}

Then, after featurization, we have for the initial node $\mathbf{s}_i'$ and edge features $e_{ij}'$:
\begin{itemize}
    \item For any node $i$, the initial node features $\mathbf{s}_i'$ is the concatenation of one-hot embedding for its amino acid type $aa_{i}$ and the ESM2 language model embedding. 
    \item For any edge $e_{i \rightarrow j}$, the initial edge features $e_{ij}'$ is defined as $e_{ij}' = \mu (r_{ij})$; Where $\mu (r_{ij})$ is the radial basis embeddings of length $r_{ij}$ and $r_{ij}$ is the norm of the displacement vector $\vec{r}_{ij} = \mathbf{x}_j - \mathbf{x}_i$. 
\end{itemize}

The initial node features $\mathbf{s}_i'$ only contains the invariant features.

\subsection{embedding module}
\label{sec:embedding_method}

After featurization, each node features are concatenated with sinusoidal embeddings of the diffusion time $\tau$. Then, the node features are embedded with shared-weights MLPs; The edge features are embedded with different MLPs depending on the type of the edges (intra- or inter- protein edges). Finally, for any node $i$ and any edge $e_{i \rightarrow j}$, we obtain the embedded node features $\mathbf{s}_i^{(0)}$ and the embedded edge features $e_{ij}^{(0)}$ from initial node features $\mathbf{s}_i'$ and the initial edge features $e_{ij}'$.

\subsection{interaction module}
\label{sec:interaction_method}

In each layer of the interaction module, for every pair of nodes within the four graphs $\{ \mathcal{G}_{rr}, \mathcal{G}_{ll}, \mathcal{G}_{rl}, \mathcal{G}_{lr} \}$, we create messages by performing tensor products of the current node features with the spherical harmonic representations of the edge vectors. The weights for this tensor product are determined by considering the edge embeddings as well as the scalar features $\mathbf{s}$ associated with both the source and destination nodes. These messages are subsequently gathered at each node and serve to update the node's current features. Then, in detail, this process is applied to each node $a$ in each intra- or inter- graphs defined previously. \\

With input node features in the below interaction model layer defined as $\mathbf{h}_i^{(0)} = \mathbf{s}_i^{(0)}$, we have, for any node $i$ in the ligand (with type $t_i = lig$) or in the receptor (with type $t_i = rec$), the aggregation of the messages happens in the intra- graph with edges $\{e_{i \rightarrow j} | j \in \mathcal{N}_i^{(t_i)}\}$ and in the inter- graph with edges $\{e_{i \rightarrow j} | j \in \mathcal{N}_i^{(\{rec, lig\} \backslash t_i)}\}$:

\begin{gather}
  \mathbf{h}_i^{(l+1)} \leftarrow \mathbf{h}_i^{(l)} \underset{t \in \{lig, rec\}}{\oplus} BN^{(t_i, t)} (\frac{1}{|\mathcal{N}_i^{(t)}|} \sum_{j \in \mathcal{N}_i^{(t)}} Y(\vec{r}_{ij}) \otimes_{\psi_{ij}} \mathbf{h}_j^{(l)}) \\
  \text{with } \psi_{ij} = \Psi^{(t_i, t)}(e_{ij}^{(0)}, \mathbf{s}_i^{(l)}, \mathbf{s}_j^{(l)}) \notag 
\end{gather}

Here, $l$ is the layer index ranging from $0$ to $L$ where $L=4$ is used in our model. $t$ represents a node type chosen arbitrarily from ligand or receptor $\{rec, lig\}$, $\mathcal{N}_i^{(t)}$ denotes the neighbors of node $i$ belonging to type $t$, $Y$ encompasses spherical harmonics up to $\ell2$ and $BN$ stands for batch normalization with equivariant properties. The output orders are limited to a maximum of $\ell1$. All trainable parameters are encompassed within $\Psi$, a collection of MLPs, each employing distinct sets of weights corresponding to different edge types. \\

\subsection{output module}
\textbf{Translational and rotational scores.} We conduct a convolution operation for each of the two outputs, where we convolve each ligand node with the unweighted center of mass c.

\begin{gather}
  \mathbf{v} \leftarrow \frac{1}{|\mathcal{V}_{lig}|} \sum_{i \in \mathcal{V}_{lig}} Y(\vec{r}_{ci}) \otimes_{\psi_{ci}} \mathbf{h}_i^{(L)} \\
  \text{with } \psi_{ci} = \Psi(\mu(r_{ci}), \mathbf{s}_i^{(L)}) \notag
\end{gather}

We constrain the output of vector $\mathbf{v}$ to include a single odd and a single even $\ell1$ vector for each of the two scores. This constraint ensures that the output remains within specific parity boundaries. Because we are employing coarse-grained representations of the protein, the resulting score does not adhere to strict even or odd properties. To reconcile this, we combine the representations of the even and odd vectors from $\mathbf{v}$ through summation. This summation is performed to capture the essential information while accommodating the non-binary nature of the score.
Subsequently, the magnitude of these vectors (while preserving their direction) is adjusted using a MLP. This MLP takes as input the current magnitude and incorporates sinusoidal embeddings of the diffusion time. Finally, to revert the previous normalization, we rescale the outputs. For the translational score, this involves multiplication by $1/\sigma_{tr}$. For the rotational score with diffusion parameter $\sigma_{rot}$, this rescaling is determined based on precomputed numerical values representing the expected magnitude of a score in SO(3).

\textbf{Intra-flexibility rotation and translation score.} We conduct additional tensor product operation like the interaction module for each residue-level rotation and translation score for the receptor and ligand respectively:

\begin{gather}
  \mathbf{h}_i^{(L+1)} \leftarrow \mathbf{h}_i^{(L)} \underset{t \in \{lig, rec\}}{\oplus}  (\frac{1}{|\mathcal{N}_i^{(t)}|} \sum_{j \in \mathcal{N}_i^{(t)}} Y(\vec{r}_{ij}) \otimes_{\psi_{ij}} \mathbf{h}_j^{(L)}) \\
  \text{with } \psi_{ij} = \Psi^{(t_i, t)}(e_{ij}^{(0)}, \mathbf{s}_i^{(L)}, \mathbf{s}_j^{(L)}) \notag 
\end{gather}

We get the $\ell1$ even and odd outputs from the resulted updated feature $\mathbf{h}_i^{(L+1)}$ for residue-level rotations and translations of receptor and ligand respectively. After the same above procedures to fix the magnitude, we can get the final score predictions.  

% every node $a$ in the complex as $h_a^{even}$ and $h_a^{odd}$. Finally, we can get, for receptor and ligand proteins, two vectors of $m$ scalars as the individual normal modes coefficient score predictions after a MLP $(\Pi)$, where $m$ is the number of non-trivial normal modes we choose to use.
% \begin{gather}
%   \mathbf{\lambda}_t = \Pi(h^{even}, h^{odd}) \\
%   \text{where } t \in \{l, r\}  
% \end{gather}

\section{Residue-Level Flow-Matching} \label{sec:appendixC}
For translations, this corresponds to learning
\[
\left[ s_{\theta}(\cdot,\tau) \right]_{\mathbf{t}_\mathrm{res}} \approx \mathbf{t}^0_\mathrm{res} - \mathbf{t}^\tau_\mathrm{res},
\]
Here, the subscript \(\mathbf{t}_\mathrm{res}\)
denotes that we are taking the translation output from the full output of \(s_{\theta}(\cdot,\tau)\).

Accordingly, the dynamics governing the evolution of translations become
\[
\frac{d\mathbf{t}_\mathrm{res}^\tau}{d\tau} = \frac{\alpha'(\tau)}{1-\alpha(\tau)}\,\left[s_{\theta}\bigl(\cdot,\tau\bigr)\right]_{\mathbf{t}_\mathrm{res}},
\]
and it analogously applies for rotations.\\
% \[
% \frac{dR^\tau_\mathrm{res}}{d\tau} = \frac{\alpha'(\tau)}{1-\alpha(\tau)}\,\left[s_{\theta}\bigl(\cdot,\tau\bigr)\right]_{R_\mathrm{res}}.
% \]

The corresponding conditional flow matching losses are defined as
\[
\begin{aligned}
L^{(\mathbf{t}_\mathrm{res})}_{\text{CFM}}(\theta) &= \mathbb{E}_{\tau\sim \mathcal{U}(0,1)}\Bigl\|\left[ s_{\theta}(\cdot,\tau) \right]_{\mathbf{t}_\mathrm{res}} - \Bigl(\mathbf{t}_\mathrm{res}^0 - \mathbf{t}_\mathrm{res}^\tau\Bigr) \Bigr\|^2,\\[1ex]
L^{(R_\mathrm{res})}_{\text{CFM}}(\theta) &= \mathbb{E}_{\tau\sim \mathcal{U}(0,1)}\Bigl\|\left[s_{\theta}(\cdot,\tau)\right]_{R_\mathrm{res}} - \operatorname{Log}\Bigl(R_\mathrm{res}^0 (R_\mathrm{res}^\tau)^{-1}\Bigr) \Bigr\|^2.
\end{aligned}
\]

These loss functions encourage the learned flow maps to accurately capture the residual corrections needed to drive an intermediate state toward the target conformation.\\

During reverse sampling, the states are evolved from \(\tau=1\) to \(\tau=0\) via Euler-like updates:
\[
\begin{aligned}
\mathbf{t}^{\tau-\Delta \tau}_\mathrm{res} &\approx \mathbf{t}_\mathrm{res}^\tau + \Delta \tau\,\frac{\alpha'(\tau)}{1-\alpha(\tau)}\,\left[ s_{\theta}(\cdot,\tau) \right]_{\mathbf{t}_\mathrm{res}},\\[1ex]
R^{\tau-\Delta \tau}_\mathrm{res} &\approx \exp\Biggl(\Delta \tau\,\frac{\alpha'(\tau)}{1-\alpha(\tau)}\,\left[ s_{\theta}(\cdot,\tau) \right]_{R_\mathrm{res}}\Biggr) \, R^\tau_\mathrm{res}.
\end{aligned}
\]

\section{Training and Hyperparameters Tuning} \label{sec:appendixD}

We use learning rate 1E-3 for the training phase on the DIPS-AF dataset and 2E-4 for the fine tuning phase on the DB5.5 dataset. For hyperparameter tuning, we tried grid search in a small scale to find the optimal loss weights. We fixed all sub-diffusion loss weights to 1.0 and conducted a grid search to optimize the weight parameters \(w_{\text{iFAPE}}\) and \(w_{\text{lddt}}\). Specifically, we explored \(w_{\text{iFAPE}}\) values within \(\{1.0, 1.5, 2.0\}\) and \(w_{\text{lddt}}\) values within \(\{0.8, 1.0, 1.2\}\). This range was chosen to ensure that all weighted sub-loss components had comparable magnitudes during training. Evaluation of the unweighted sub-losses on the validation set across different trained models revealed negligible performance differences among the parameter combinations tested. Consequently, we selected \(w_{\text{iFAPE}} = 1.0\) and \(w_{\text{lddt}} = 1.0\) as optimal weights for our final model.

Each of our models is pretrained on the DIPS-AF dataset for around 7 days using 8 A100 GPUs and finetuned on the DB5.5 dataset for about 10 hours. For the finetuned confidence models in Section 4.4, the finetuning process takes approximately 2 days with the same GPU configuration.

\section{Supplementary Tables and Figures} \label{sec:appendixE}

% \subsection{Ablation study results stratified by difficulty level}

\begin{table}[h]
\caption{\textbf{Detailed statistics of the DIPS-AF dataset across splits.} The docking difficulty characterization follows the same criteria used for the DB5.5 dataset according to fnonnat and iRMSD calculated between superimposed unbound structures and bound structures.}
\centering
\begin{tabular}{lccc}
\hline
\textbf{Split} & \textbf{Rigid-body} & \textbf{Medium} & \textbf{Difficult} \\
\hline
Train & 29,260 & 3,161 & 4,639 \\
Validation & 778 & 43 & 98 \\
Test & 685 & 116 & 107 \\
\hline
\textbf{Total} & \textbf{30,723} & \textbf{3,320} & \textbf{4,844} \\
\hline
\end{tabular}
\label{tab:dips_af_detailed_stats}
\end{table}

The Table~\ref{tab:dips_af_detailed_stats} reveals that the DIPS-AF dataset is heavily skewed toward rigid-body docking cases. Also notably, the unbound structures in DIPS-AF are generated using AlphaFold2 model, which is primarily trained on native complexes. As a result, these AF-predicted "unbound" structures often inherit bound-like features, reducing their conformational flexibility. This implicit bias makes the DIPS-AF dataset effectively a rigid-body dominated dataset. Consequently, evaluation on the DB5.5 benchmark, which features greater conformational variability and differs substantially in distribution, requires the model to overcome a significant domain shift.//

\begin{table}[htbp]
    \setlength{\tabcolsep}{1pt} % Adjust column spacing
    \caption{\textbf{DB5.5 Test Set Evaluation Top-1 Results By Difficulty Level: Ablation Study.} Bold numbers indicate the lowest (best) mean RMSD values, while dagger-marked numbers represent the second lowest values in each column.}
    \label{tab:db5_ablation_study_difficulty_level}
    \begin{tabular}{lccc|ccc}
    \toprule
    & \multicolumn{3}{c}{Complex RMSD Mean$\pm$Std (Å)} & \multicolumn{3}{c}{Interface RMSD Mean$\pm$Std (Å)} \\
    \cmidrule(lr){2-4} \cmidrule(lr){5-7} 
     & rigid-body  & medium  & difficult  & rigid-body  & medium  & difficult  \\
    \textbf{Methods} & (16 cases)  & (6 cases)  & (3 cases)  & (16 cases)  & (6 cases)  & (3 cases)   \\
    \midrule
    Model0              & 15.60$\pm$5.3  & $16.54^{\dagger}$$\pm$5.3  & 19.17$\pm$3.9  & 13.65$\pm$4.9  & $13.04^{\dagger}$$\pm$3.9  & $15.93^{\dagger}$$\pm$5.1  \\
    Model\_NMFeatures    & $15.42^{\dagger}$$\pm$5.6  & 18.60$\pm$6.2  & 19.72$\pm$4.6  & $13.15^{\dagger}$$\pm$4.6  & 15.89$\pm$5.7  & 16.09$\pm$1.8  \\
    Model\_adaptive      & \textbf{14.18}$\pm$4.0  & \textbf{12.49}$\pm$3.9  & \textbf{16.99}$\pm$4.1  & \textbf{12.86}$\pm$5.2  & \textbf{11.81}$\pm$4.5  & \textbf{12.26}$\pm$2.9  \\
    (Model\_adaptive\_noPretrain) & 16.86$\pm$4.8  & 17.37$\pm$4.8  & $18.95^{\dagger}$$\pm$3.9  & 14.95$\pm$5.6  & 13.70$\pm$4.2  & 16.78$\pm$5.1  \\
    % (Ensemble\_iRMSDheadModel\_l2RegModel)  &   14.61$\pm$5.3 & 14.46$\pm$5.2 & 21.26$\pm$4.9 & 13.32$\pm$5.5 & 13.76$\pm$5.7 & 15.95$\pm$3.0 \\
    \bottomrule
    \end{tabular}
\end{table}

The ablation study presented in Table~\ref{tab:db5_ablation_study_difficulty_level} evaluates the incremental impact of integrating normal mode (NM) features and an adaptive training schedule, as well as the effect of pretraining, on the prediction accuracy measured by complex and interface RMSD across difficulty levels.

Model0 serves as a baseline, showing intermediate performance, particularly notable on medium-difficulty cases. Upon introducing NM features alone (Mode\_NMFeatures), the performance slightly improves for rigid-body cases (second-best), yet noticeably deteriorates for medium and difficult scenarios. This indicates that the standalone incorporation of NM features does not consistently enhance docking performance, particularly for more challenging cases.

The integration of an adaptive training strategy (Model\_adaptive) results in substantial improvements across all levels of difficulty, achieving the best RMSD scores overall. This highlights that dynamically adjusting training conditions significantly enhances the model's generalization capabilities and accuracy in predicting realistic docking poses. Notably, this improvement suggests a synergistic benefit, wherein the NM features, previously marginally effective, become notably beneficial when employed within an adaptive training framework. Therefore, NM features provide valuable complementary structural information, whose full potential is realized under adaptive training conditions.

Further comparison with a non-pretrained variant (Model\_adaptive\_noPretrain) reveals significant performance drops, particularly evident for rigid-body and medium cases, underscoring the importance of pretrained weights. These results highlight that pretrained initialization confers a beneficial prior, critical for achieving lower RMSD values, particularly in more easily docked cases, indicating that pretrained knowledge is essential for the model’s robustness.

\clearpage
% \section{Appendix}
\newpage
% \bibliography{ref}

\bibliographystyle{plainnat}

\end{document}